\documentclass{emulateapj}
\shorttitle{High dynamic range Monte Carlo}
\shortauthors{C.W. Ormel \& M. Spaans}
\usepackage[varg]{txfonts}
\newcommand{\se}[1]{\mbox{\S\ \ref{sec:#1}}}

\newcommand{\eq}[1]{\mbox{equation\ (\ref{eq:#1})}}
\newcommand{\eqp}[1]{\mbox{eq.\ [\ref{eq:#1}]}}
\newcommand{\Eq}[1]{\mbox{Equation\ (\ref{eq:#1})}}
\newcommand{\fg}[1]{\mbox{Fig.\ \ref{fig:#1}}}
\newcommand{\Fg}[1]{\mbox{Figure\ \ref{fig:#1}}}
\newcommand{\Tb}[1]{\mbox{Table\ \ref{tab:#1}}}
\newcommand{\ie}{i.e.,}
\newcommand{\eg}{e.g.,}
\newcommand{\sumi}{(i)}
\newcommand{\sumii}{(ii)}
\newcommand{\sumiii}{(iii)}

\begin{abstract}
We present a method which extends Monte Carlo studies to situations that require a large dynamic range in particle number. The underlying idea is that, in order to calculate the collisional evolution of a system, some particle interactions are more important than others and require more resolution, while the behavior of the less important, usually of smaller mass, particles can be considered collectively. In this approximation groups of identical particles, sharing the same mass and structural parameters, operate as one unit. The amount of grouping is determined by the zoom factor -- a free parameter that determines on which particles the computational effort is focused. Two methods for choosing the zoom factors are discussed: the `equal mass method,' in which the groups trace the mass density of the distribution, and the `distribution method,' which additionally follows fluctuations in the distribution. Both methods achieve excellent correspondence with analytic solutions to the Smoluchowski coagulation equation. The grouping method is furthermore applied to simulations involving runaway kernels, where the particle interaction rate is a strong function of particle mass, and to situations that include catastrophic fragmentation.  For the runaway simulations previous predictions for the decrease of the runaway timescale with the initial number of particles ${\cal N}$ are reconfirmed, extending ${\cal N}$ to $10^{160}$. Astrophysical applications include modeling of dust coagulation, planetesimal accretion, and the dynamical evolution of stars in large globular clusters. The proposed method is a powerful tool to compute the evolution of any system where the particles interact through discrete events, with the particle properties characterized by structural parameters.
\end{abstract}
\begin{document}
\title{Monte Carlo simulation of particle interactions at high dynamic range: Advancing beyond the Googol}
\author{C.W. Ormel and M. Spaans}
\affil{Kapteyn Astronomical Institute, University of Groningen\\
       P.O. Box 800, 9700 AV Groningen,\\ 
       The Netherlands}
\email{ormel@astro.rug.nl, spaans@astro.rug.nl}
\section{Introduction}
\label{sec:intro}
In the field of natural sciences the dynamical evolution of a population is often determined by interactions that operate at short range, \eg\ when the members come into contact, but which do otherwise not influence each other. The dynamical evolution of the system is then regulated by discrete events and the system can be described statistically. The most direct interaction is, of course, the merging of two bodies that come into contact. Examples are numerous. In the Earth's atmosphere water vapour condenses on cloud droplets (tiny water drops), which merge to form rain drops. In biology, the size distribution of many species of animals can be found by assuming animal groups merge when they make contact. The distribution is then determined by a power-law, which seems to agree with observations of, \eg\ schools of tuna fish or herds of African buffaloes \citep{1999PNAS...96.4472B}. Similarly, a power-law emerges when bigger organisms are assumed to feed on smaller ones \citep{2001EL.....55..774C}. In astrophysics, too, coagulation processes appear. The transition of the dust component in protoplanetary disks -- from (sub)micron grains to planets -- covers a huge dynamic range: many tens of orders of magnitude in mass. Due to small molecular forces, micron-sized grains will stick when they meet \citep{2004ASPC..309..369B}. How sticking continues as particles grow to macroscopic proportions ($\sim$cm or larger) is still unclear but once the km-size is reached gravity will dominate the accretion process \citep{1993ARA&A..31..129L,2004ARA&A..42..549G}. Another example where coagulation is of importance are dense stellar systems. Although stellar densities are usually too low for collisions to be likely on a Hubble timescale, it has been shown that mechanisms, \eg\ mass-segregation or equipartition, conspire to make stellar collisions feasible \citep{2006MNRAS.368..121F}.

The coagulation process, excluding fragmentation and injection, is described mathematically by the Smoluchowski equation \citep{1916ZPhy...17..557S}, which in the continuous limit reads
\begin{eqnarray}
  \frac{\partial f(m)}{\partial t} = - f(m) \int \mathrm{d}m'\ f(m') K(m,m') \nonumber \\
                                    + \frac{1}{2} \int \mathrm{d}m'\ f(m-m')f(m') K(m-m',m),
  \label{eq:smoluchowski}
\end{eqnarray}
where the terms on the RHS denote, respectively, the loss and gain terms of particles of mass $m$. The distribution function, $f(m)$, gives the number density of particles (\eg\ dust grains, stars, etc.) of mass $m$; \ie
\begin{eqnarray}
  f(m) \cdot \mathrm{d}m = \textrm{number density of particles in} \nonumber \\
                      \textrm{mass-interval ($m, m+dm$)}.
  \label{eq:distr}
\end{eqnarray}
Given that the initial conditions of the distribution, \ie\ $f(m,t=0)$ are specified, the coagulation is fully determined by the coagulation kernel, $K(m,m')$, which is the product of the cross section of collision and the relative velocity of the particles, $K(m,m') = \sigma(m,m') \Delta v(m,m')$. It gives the rate coefficient (units: $\mathrm{cm^3\ s^{-1}}$) for the collision between two particles of mass $m$ and $m'$. For only three distinct cases of $K$ does \eq{smoluchowski} have a, closed form, analytic solution \citep{1994Icar..107..404T}. These are:

\begin{itemize}
  \setlength{\itemsep}{-1mm}
    \item The constant kernel, $K = \textrm{cnst.}$;
    \item The additative (or sum) kernel, $K \propto (m_i + m_j)$;
    \item The product kernel, $K \propto m_i m_j$;
\end{itemize}
\begin{deluxetable}{lllll}
  \tablecaption{\label{tab:moments} Moments}
  \tablehead{ \colhead{kernel}          & \colhead{$K_{ij}$}                & \colhead{$M_0(t)$}                  & \colhead{$M_1(t)$}    & \colhead{$M_2(t)$} }
  \startdata
  constant        & $1$                     & $(1+\case{1}{2}t)^{-1}$  & $1$         & $1+t$ \\
  sum             & $\case{1}{2}(m_i+m_j)$  & $\exp[-\case{1}{2}t]$    & $1$         & $\exp[t]$\\
  product         & $m_i m_j$               & $1-\case{1}{2}t$         & $1$         & $(1-t)^{-1}$
  \enddata
  \tablecomments{Moment expressions for the classical analytic kernels: $M_0$ describes the particle density, $M_1$ the mass density (which is conserved) and $M_2$ is related to the mass peak, $m_p$, of the distribution. The initial distribution at $t=0$ is monodisperse, $f(m,t=0) = \delta(m-1)$. The numerical factor of $1/2$ in the sum kernel causes its evolution to coincide with the other kernels at small $t$. The expressions for the product kernel, $K=m_im_j$ become invalid after the formation of the gel, at $t=1$. The mean mass $\langle m \rangle$ and peak mass $m_p$ are related to the moments as $\langle m \rangle = M_1/M_0$ and $m_p=M_2/M_1$.}
\end{deluxetable}
These three kernels each show their own characteristic behavior. This can be seen by considering the moments of the distribution
\begin{equation}
  M_p(t) = \int_0^\infty m^p f(m,t)\ \mathrm{d}m.
  \label{eq:momentdef}
\end{equation}
Multiplying the Smoluchowski equation by $m^p$ and integrating over $m$ yields the moment equation \citep{2003PhR...383...95L}
\begin{eqnarray}
  \frac{\mathrm{d}M_p}{\mathrm{d}t} = \frac{1}{2} \int_0^\infty \mathrm{d}m\ \mathrm{d}m'\ K(m,m') f(m,t) f(m',t) \nonumber \\
   \times \left[ (m+m')^p - m^p - m'^p \right],
  \label{eq:momentev}
\end{eqnarray}
from which $M_p(t)$ can be found. In \Tb{moments} the zeroth, first and second moments as function of time for monodisperse initial conditions are given for the three classes of kernels. The first moment $M_1$, the mass density, is conserved. The zeroth moment, $M_0$, describes the (decreasing) particle density; it therefore defines the mean mass of the population, $\langle m \rangle = M_1/M_0$. The second moment, $M_2$, is of considerable importance in this paper as it traces the mass distribution on a logarithmic scale, \ie\ $mf(m)\mathrm{d}m = m^2f(m)\mathrm{d}\log m$. We will therefore define the peak mass as $m_p = M_2(t)/M_1(t)$.
For most continuous distributions $m_p$ corresponds to the particle sizes that (together) dominate the total mass of the population. 

These three kernels are each representative of a very distinct growth behavior. In the constant kernel growth is orderly: not only do $\langle m \rangle$ and $m_p$ evolve linearly with time, their ratio, which indicates the (relative) spread in the distribution, stays constant as well. This relative narrowness of the distribution means that this kernel can be relatively easily simulated by numerical methods. The situation for the sum kernel, however, is different. Initially, ($t\ll1$) it evolves similarly as the constant kernel, but after $t\sim1$ it starts to grow exponentially. The exponential nature of the growth not only holds for $\langle m \rangle$ or $m_p$ but also for the spread in the distribution, $m_p/\langle m \rangle \sim \exp[\case{1}{2}t]$ (see \fg{distrib}). 

The third class of kernels are the runaway kernels, of which $K_{ij}=m_im_j$ is its analytic representative. From \Tb{moments} a peculiarity can be noted: the moment expressions do not make sense after a certain amount of time: the mean density will become negative after $t=2$, whereas $M_2(t)$ becomes infinite at $t=1$. The reason for this behavior is that at $t=1$ a runaway particle of mass $m_R(t)$ is created that separates from the rest of the distribution. In theoretical studies of coagulation this particle is called the `gel' and $m_R(t)$ becomes infinite at precisely $t=1$. In physical situations, where the total mass in the system is finite, $m_R(t)$ grows continuously with time but nevertheless separates from the distribution. This discontinuity in the distribution is also the reason why the moment expressions become invalid after $t=1$.

In systems where small number statistics (one runaway particle) become important the assumptions underlying the Smoluchowski equation become invalid. Coagulation is more accurately described by the stochastic coagulation equation which assigns a probability that the system is in a specific state $\{n_k\}(t)$, with $\{n_k\}$ the discrete particle numbers for the mass bins $k$. The Smoluchowski equation only follows when certain assumptions are made, like $\langle n_i n_j \rangle = \langle n_i \rangle \langle n_j \rangle$, which break down when $i$ and $j$ are statistically correlated \citep{1975JAtS...32.1977G}. Clearly \eq{smoluchowski} is dangerous when it is used to describe the coagulation process -- a process that is inherently discrete.

In many situations of physical interest $K=\sigma \Delta v$ cannot be expressed in the simple forms above where analytic solutions are available. If $K$ can be expressed in terms of a simple mass-scaling behavior of the form $K \propto \textrm{(mass)}^\beta$, then semi-analytic solutions are available \citep{1979ApJ...229..242S}. However, in many physical situations $\beta$ will not  be constant over the whole mass range under consideration. 
For example, in studies of dust coagulation in protoplanetary disks, relative velocities can be divided into several `regimes,' each with a different velocity behavior \citep[\eg][]{2007A&A...466..413O}. Another example, is the enhancement of the collisional cross section by gravitational focussing, \ie 
\begin{equation}
  \sigma = \pi R_s^2 \left[ 1 + \frac{2G(m_i+m_j)}{R_s (\Delta v_{ij})^2} \right],
  \label{eq:gravfoc}
\end{equation}
where $R_s$ is the sum of the radii of the bodies and $G$ Newton's gravitational constant. \Eq{gravfoc} displays a natural break in the scaling of $\sigma$ with mass: \ie\ $\sigma \propto m^{2/3}$ at low masses vs.\ $\sigma \propto m^{4/3}$ at high masses. The break happens at the point where the relative velocity of the particles starts to become less than the escape velocity $v_\mathrm{esc} = \sqrt{2G(m_i+m_j)/R_s}$ of the biggest body, $\Delta v_{ij}/v_\mathrm{esc}<1$.

So in most situations a numerical approach to the coagulation problem is required. Most directly, \eq{smoluchowski} can be integrated numerically. This entails dividing up the mass-axis into discrete bins and applying \eq{smoluchowski} to all of these bins. However, this is a rather elaborate procedure as, \eg\ mass conservation is not intrinsically conserved, and most works have actually avoided to explicitly use the Smoluchowski equation.
For example, \citet{1991Icar...92..147S} and \citet{1999EP&S...51..205I} use a (fixed) binning method that assumes the mass inside a bin is distributed as a power-law; coagulation is then an interchange of mass between the various bins. Recently, \citet{2007arXiv0711.2192B} have modified a matrix method, the Podolak algorithm, (see also \citealt{1969JAtS...26.1060K}) which distributes the mass of a coagulation event evenly between its two nearest grid points. \citet{1990Icar...88..336W} and \citet{1999EP&S...51..205I} use a method where the mass is divided over a discrete number of `batches,' each characterized by a single mass. Interactions with other batches and with themselves cause the batches to grow, \ie\ move up the mass scale. This representation of intrinsically discrete quantities is especially useful in the case of non-continuous distributions (runaway kernels). 

Bin size techniques are generally one-dimensional, \ie\ the quantities that enter \eq{smoluchowski} (cross sections, velocities) depend on mass only. This approximation may neglect \textit{variables} that play a pivotal role in the coagulation process. For example, a charge distribution on grains \citep{2001LPI....32.1262M,2005NJPh....7..227K} is difficult to include in this way. A more subtle example concerns the coagulation of dust particles in disks. Here, collisions affect the internal structure of the particles such that their porosity is not constant, which affects their aerodynamic properties and hence their collisional evolution. \citet{1993A&A...280..617O} has extended the Smoluchowski formalism to solve the two-dimensional distribution function for the collisional evolution of dust aggregates, using their fluffiness as the second variable. However, it is clear that the increased number of connections between the bins, caused by the higher dimensionality of the problem, makes the procedure more elaborate, and slower -- impractical, perhaps, for three or more dimensions. 

\begin{figure}[t]
  \plotone{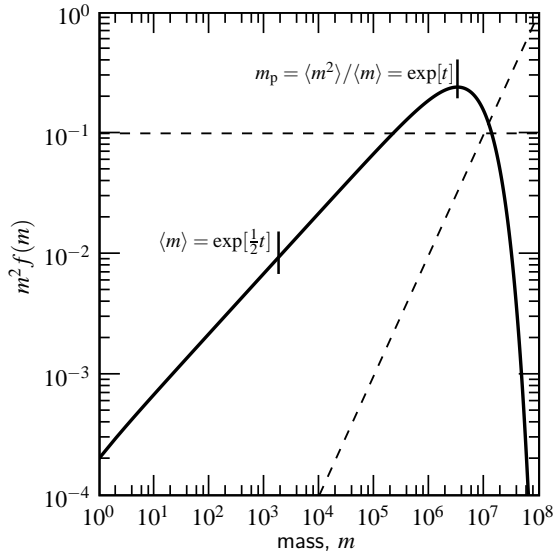}
  \caption{\label{fig:distrib}The $m^2 f(m)$ mass function of the sum kernel ($K_{ij} = \case{1}{2}(m_i+m_j)$) at $t=15$ (\textit{solid curve}). At this time the mass function has achieved self-similarity. The mean mass, $\langle m \rangle$, and peak mass, $m_p$, are indicated. Dashed lines of logarithmic slopes 0 and 1, respectively, trace areas of equal mass- and number density per logarithmic unit of mass. Due to the $f(m) \propto m^{-3/2}$ fall-off towards small masses, the small particles dominate by number, yet most of the mass resides in the large particles (around $m=m_p$).}
\end{figure}
In these cases one has to resort to Monte Carlo (MC) simulations. A good description of how a direct-simulation MC method is implemented is outlined by \citet{1975JAtS...32.1977G} which we will briefly summarize below (\se{code}). In MC methods $K(m,m')$ determines the probability an event (here: the collision between the particles) takes place. For multi-dimensional models it is probably the preferred method. Besides, the MC method as described by \citet{1975JAtS...32.1977G} provides an exact description of the stochastic coagulation process, and will also accurately describe runaway kernels.  However, in MC simulations one is limited to the finite number of particles that can be followed. It is therefore best suited for distributions that remain relatively narrow, such that a modest number of particles gives a good knowledge of the distribution function $f$.  On the other hand, when the distribution is wide a large number of particles must be present in order to obtain a good characterization of $f$; the required particle numbers then severely limit the applicability of the code.

These effects can best be illustrated in the case of the sum kernel. In \fg{distrib} we have plotted the theoretical value of $m^2 f(m)$ at $t=15$ \citep{1994Icar..107..404T} corresponding to $K_{ij} = \case{1}{2} (m_i + m_j)$, and an initial monodisperse distribution of mass $m_0=1$ and $f(m,t=0) = \delta(m-m_0)$. At $t=15$ the distribution has reached self-similarity and the shape is the characteristic one for the sum kernel. In \fg{distrib} $m^2f(m)$ is plotted showing (per logarithmic mass interval) the mass density distribution. Due to the $m^{-3/2}$ fall-off of $f(m)$ with mass in the low-$m$ tail of the distribution, the smaller ($m \sim m_0$) particles dominate by number, although it is the larger particles that contain most of the mass. If, for example, the total simulated mass is $M_\mathrm{tot}$, there are $\sim M_\mathrm{tot}/m_p$ particles representing the mass peak, and $\sim M_\mathrm{tot} /\langle m \rangle$ particles in total. The fraction of $m_p$ particles then decreases with increasing $m_p$ as $\langle m \rangle/m_p \propto m_p^{-1/2}$. This behavior inevitably causes computational problems since a few massive particles remain in comparison to a huge number of small particles, while both regimes must be included in order to obtain a good characterization of $f(m)$. With increasing $m_p$, a point is reached at which there are simply not enough particles to resolve the complete distribution. 

In short, the dynamic range for these kinds of simulations is too large to tackle with conventional MC-methods; more particles are required but this is impossible due to computational constraints. To overcome these problems -- \ie\ to extend MC-methods to situations that require high dynamic range -- we will introduce the assumption that one simulation particle can represent multiple physical particles: the grouping method. \se{group2} introduces the grouping method. First the framework of a MC simulation is described in \se{code}, then \se{group2}-\se{magdemag} outline the grouping method in detail. In \se{results} the new method is tested against the analytic solution of the sum kernel (\se{sum}) and the product kernel (\se{product}). In \se{runaway}, we consider a class of runaway kernels and test the assumption that the runaway timescale goes down with the size of the simulation (keeping the physical quantities such as the initial mass density constant). Next, in \se{frag} we present steady-state distributions that result from fragmentation of the largest particles. We briefly summarize our findings and discuss astrophysical applications in \se{summ}.

\section{The grouping algorithm}
\label{sec:group}
Before introducing the grouping method (\se{group2}) we first briefly outline the main features of a direct simulation Monte Carlo code.
\subsection{\label{sec:code}The Monte Carlo code}
\noindent
\begin{deluxetable}{lp{7cm}}
  \tablecaption{\label{tab:def}List of symbols}
  \tablehead{ \colhead{symbol} & \colhead{description} }
  \startdata
    $\beta$       & mass dependence of kernels as in $K\propto m^\beta$\\
    $\Delta m_i$    & mass change of $i$-particle due to accretion of $j$-particles \\
    $\lambda_{ij}$  & collision rate [$\mathrm{s}^{-1}$] between species $i$ and $j$ (including duplicates) \\
    $C_{ij}$        & collision rate [$\mathrm{s}^{-1}$] between particle $i$ and $j$ (excluding duplicates) \\
    $K_{ij}$        & coagulation kernel or rate coefficient [$\mathrm{cm^3\ s^{-1}}$] (velocity $\times$ cross-section) \\
    $M_k$           & $k$-th moment of distribution\\
    $M_{L1}, M_{L2}$& first, second most massive particle\\
    $N_\epsilon$    & group splitting factor \\
    $N_g$  & number of groups, $\sum_{i=1}^{N_s} w_i$ \\
    $N_p$  & number of total particles in simulations \\
    $N_s$  & number of species \\
    $N_x^\ast$      & target number of species ($N_s^\ast$) or groups ($N_g^\ast$)\\
    ${\cal N}$      & initial number of particles, $N_p(0)$ \\
    ${\cal V}$      & simulation volume \\
    $f_\epsilon$    & maximum allowed fractional mass change of a particle during group collisions\\
    $g_i$           & occupancy number of species $i$; number of duplicates \\
    $m_\star$       & characteristic mass of distribution (determining $\{z_i\}$ in equal-mass method)\\
    $\langle m \rangle$  & mean mass of the distribution $M_1/M_0$\\
    $m_i$           & mass of species $i$\\
    $m_p$           & peak mass of the distribution $M_2/M_1$\\
    $w_i$           & group number; number of groups for species $i$\\
    $z_i$           & zoom number of species $i$
  \enddata
\tablecomments{Definitions for key parameters used in this paper.}
\end{deluxetable}
The direct simulation Monte Carlo method in its simplest form consists of three steps \citep{1975JAtS...32.1977G}:
\begin{itemize}
\setlength{\itemsep}{-1mm}
\item[\sumi] calculation/update of the collision rates, $\{C_{ij}\}$;
\item[\sumii] selection the collision by random numbers: which particles collide and the timestep involved;
\item[\sumiii] execution of the collision: removal of the collision partners and addition of the new particle(s).
\end{itemize}
A Monte Carlo simulation starts with the calculation of the collision rates between the particles, $C_{ij} = K_{ij}/{\cal V}$ where ${\cal V}$ is the simulated volume and $1 \le i,j \le N_p$. Here, $N_p$ is the number of particles in the simulation and $C_{ij}$, the collision rate, is related to the probability of collision, such that $C_{ij} \Delta t$ is the probability that particles $i$ and $j$ collide in the next infinitesimal time $\Delta t$. In total there are $N_p(N_p-1)/2$ collision rates. A trick which saves memory requirements (although it increases CPU-time) is to store only the partial rates, $C_i = \sum_j^{N_p} C_{ij}$. The total collision rate is $C_\mathrm{tot} = \sum_i^{N_p} C_i$. In step \sumii\ three random deviates ($\overline{r}$) determine the time interval to the next collision, and the particles involved ($i$ and $j$). As described by \citet{1975JAtS...32.1977G} the time increment $\Delta t$ is found from the total collision rate, $\Delta t = - C_\mathrm{tot}^{-1} \ln \overline{r}$, whereas the particles that are involved are found from sampling the $\{C_i\}$ and $\{C_{ij}\}$. Step \sumiii\ computes the new properties of the particles that result from the collision.  The fact that in MC-methods these steps are separated shows its advantages when introducing multiple variables; by including multiple variables, the outline of the program, \ie\ steps \sumi--\sumiii, will remain the same. Finally, the $(N_p-1)$ collision rates associated with the newly created particle need to be re-calculated (step \sumi), after which the $\{C_i\}$ are updated. The steps then repeat themselves in a new cycle.

\subsubsection{Constant-$V$ vs constant-$N$ simulations}
Because each collision that results in coagulation reduces the particle number by one, all particles will have coalesced into one body after $N_p$ steps. This is the constant-${\cal V}$ algorithm: the simulated volume ${\cal V}$ and total mass stays the same. However, in many cases this is not the intended behavior: the volume ${\cal V}$ occupied by the initial number of particles in the simulation, ${\cal N} = N_p(0)$, is only a small sample of the total volume under consideration. Although ${\cal V}$ may start out as a volume large enough to represent the particle population, after sufficient coagulation (and decreasing particle numbers) it looses this property: a larger volume is needed in order to get the desired sampling. However, increasing the initial number of particles ${\cal N}$ is no solution since the number of collision rates becomes quickly impractical.

A natural solution to prevent the statistical deterioration due to the decreasing particle numbers is to keep the numbers constant: the constant-$N$ algorithm. Here, the number of (simulated) particles is artificially kept constant by \textit{duplicating} a particle after each coagulation event. Thus, mass is added to the system, which physically translates into an expansion of the simulated volume, because the volume-density stays the same. As described in \citet{1998...53..1777} the new particle is taken randomly from the existing (simulated) distribution and is therefore called a duplicate. \citet{1998...53..1777} have shown that the error introduced due to the duplication scales logarithmically with $\langle m \rangle (t)/m_0$, which is a measure of the amount of growth. This is much improved compared to the constant-${\cal V}$ method in which the reduced number of particles causes the error to scale as the square-root of the growth. In principle, this method allows for indefinite growth. Fundamentally, the assumption is that the simulated volume (containing $N_p$ particles) at all times stays representative of the parent distribution (containing $\gg N_p$ particles). We will test whether the \citet{1998...53..1777} duplication mechanism also holds in systems where the growth occurs over many orders of magnitude. 

On the other hand, in situations where runaway growth occurs, the collisional evolution will depend on the initial number of particles present in the system, even if locally the same physical conditions apply, \ie\ for the same density \citep{1990Icar...88..336W,2001Icar..150..314M}. Clearly, a constant-$N$ simulation is not the proper way to treat these kernels; we will always use a constant-${\cal V}$ algorithm when treating these kernels. It is challenging to test simulations in which ${\cal N} = N_p(0)$ becomes truly astronomical. Here, too, the grouping method will be applied.

\subsubsection{\label{sec:species}The species formalism}
A natural way to deal with the existence of a large number of \textit{identical} particles, \eg\ resulting from monodisperse initial conditions or through the duplication mechanism in constant-$N$ simulations, is to combine them in the calculation of the collision rates \citep{spouge1985mcr,2001Icar..150..314M,2002JCoPh.177..418L,2003PhRvE..67e1103L}. Since the collision rates of the duplicates are the same, it is efficient to introduce a multiplicity array, $g$, for all \textit{distinct} particles. This is like saying we have 1 particle of type 1, 5 particles of type 2 \dots $g_i$ of \textit{species} $i$. It causes the number of distinct simulation particles ($N_s$, or number of species) to become less than the total number of physical particles in the simulation ($N_p$) since 
\begin{equation}
  N_p = \sum_{i=1}^{N_s} g_i.
  \label{eq:NsNp}
\end{equation}
This algorithm is much more efficient, for it is the number of distinct particles ($N_s$) that matters in the calculation of the collision rates. The collision rate for one collision between species $i$ and $j$ then becomes
\begin{equation}
    \lambda_{ij} = g_i g_j C_{ij} \qquad (i\ne j);\qquad \lambda_{ii} = \case{1}{2}g_i (g_i-1) C_{ii} \qquad (i=j),
    \label{eq:cr1}
\end{equation}
where the $C_{ij}$ are the individual collision rates between two single particles. So the $\lambda_{ij}$ give the probability that a (single) particle of species $i$ collides with one of species $j$. After the collision partners are determined their multiplicity is first decreased ($g_i\rightarrow g_i-1$; $g_j\rightarrow g_j-1$). If this means that $g_i$ becomes zero, no particles of type $i$ exist anymore and it is consequently removed ($N_s\rightarrow N_s-1$). The new particles that result out of the collision are added as a new (and distinct) species ($N_s\rightarrow N_s+1$). When particles are characterized by only one parameter (mass) it would in this case be more efficient to assess first whether the species is really unique, but in this study we treat a general approach. Duplication causes the occupancy number of the duplicated species $k$ to increase by 1, $g_k\rightarrow g_k+1$. \citet{2002JCoPh.177..418L} describe in-depth how the species formalism can be implemented and quantify its computational efficiency. 

We note that this grouping of the collision rates is nothing more than a computational trick, which, within the assumptions of, \eg\ the duplication mechanism does not approximate the MC-method. Occupation numbers may become very large, though. For example, for monodisperse initial conditions $g_1 = N_p$ initially but $g_1$ may increase if the constant-$N$ algorithm is enforced and ${\cal V}$ grows, as in the case of the sum kernel (\fg{distrib}). In fact, the existence of a large number of identical particles is the basis of the (physical) grouping algorithm.

\subsubsection{\label{sec:Ns}Limiting $N_s$.}
The accuracy and efficiency of Monte Carlo simulations is then determined by a single parameter, $N_s$. Ideally, one wants to constrain $N_s$. In the constant-$N_s$ simulations this means that $N_s$ should fluctuate around a (fixed) target value $N_s^*$. (There is no need to enforce an exact 1-1 correspondence.) Here we list a few mechanisms that can be applied to balance $N_s$. Operations that increase $N_s$:
\begin{itemize}
\setlength{\itemsep}{-1mm}
  \item Duplication ($g_k\rightarrow g_k+1$). Although duplication does not directly increase the number of species, the increased multiplicity of a species will decrease the likelihood that it becomes empty of particles ($g_k=0$) and that it is removed;
  \item Coagulation/Fragmentation. Because we generally assume that the collision products are different species, the number of species always increases with coagualation and fragmentation. Especially fragmentation may lead to a proliferation of the number of species.
\end{itemize}
And operations that decrease the $N_s$ parameter:
\begin{itemize}
\setlength{\itemsep}{-1mm}
  \item (Physical) grouping (discussed below). This has the effect that many particles are involved in the collision and therefore increases the likelihood of a disappearing species ($g_i\rightarrow 0$).
  \item Coagulation. Requires two particles, with which again a species may disappear.
  \item Particle removal. This is the reverse of duplication. If the number of species keeps increasing, we may randomly remove particles to enhance the probability of obtaining $g_i=0$.
  \item Merging. This is the most drastic approach, in which different species are merged together in a single species whose properties reflect the properties of its progenitors. Because of the averaging, this procedure is somewhat against the spirit of the MC-methods. It is in fact a smoothening where particles of different properties are averaged, including their structural parameters. However, the mean parameters may still be applicable to describe the present state.
\end{itemize}
In this work the species-averaging is performed as follows. If a species needs to be merged, we first look for its neighbor species in terms of mass, and then mass-average over their (structural) properties. As in this work mass is the only property, it simply means that for species $i$ and $j$ their merged occupation number becomes $g=g_i+g_j$ and the new mass $(g_i m_i+g_j m_j)/(g_i+g_j)$. (If needed, the latter quantity is rounded to integer values). If structural parameters would be present, however, this procedure can be simply extended to any structural property $\psi$, \ie\ $\psi \rightarrow (g_i\psi_i + g_j\psi_j)/(g_i+g_j)$, where $\psi$ may, \eg\ represent charges or filling factor.

\subsection{\label{sec:group2}The grouping algorithm}
To simulate a coagulation process by Monte Carlo methods in a way to obtain growth over many orders of magnitude, two criteria have to be met:
\begin{itemize}
\setlength{\itemsep}{-1mm}
    \item[I] The simulation must involve a large number of physical particles $(N_p)$, such that the population is a proper sample of the continuous distribution. Physically, it means the spatial volume that is sampled will be large enough to account for all various kinds of particles.
    \item[II] The number of species ($N_s$) cannot become too large: since there are $\sim N_s^2$ collision rates to be taken care of, the computational speed will slow down rapidly with increasing $N_s$.
\end{itemize} 
To meet these conditions -- high $N_p$ and low $N_s$ -- the occupation numbers $g_i$ must be very large. This can, in principle, be achieved by duplicating particles many times. However, a new problem then arises. Since the total collision rate, $C_\mathrm{tot}$, scales with the total number of physical particles squared, \ie\ $C_\mathrm{tot} \sim N_p^{2}$, the time step per collision is proportional to $N_p^{-2}$. The simulations then seem to freeze as the required CPU-time per unit simulation time blows up. In the MC-simulation the focus lies on the particles that dominate by number, which may sometimes be irrelevant from a physical point of view. This occurs, for example, with the sum kernel. Here, the most dominant collisional process -- a collision between a small particle and a very massive one -- is rather meaningless: a single collision neither affects the mass of the massive particle nor the number of small particles. The ubiquity of many of these subleading collisions then overwhelms the more important ones involving more massive particles, a process that together is very CPU intensive. This leads to the conclusion that, given (I) and (II), a third condition must apply
\begin{itemize}
\setlength{\itemsep}{-1mm}
  \item[III] A simulation involving a large number of particles ($N_p$) yet with only 2 physical particles colliding per MC cycle can become very inefficient. Therefore, (small) particles should be grouped together -- \ie\ considered as a single unit -- during the collision.
\end{itemize}

Due to the duplication mechanism there is a natural way to group particles together. In \eq{cr1} it was already seen that identical particles can be effectively absorbed in the calculation of the collision rates. We will now take this idea further and write the $g_i$'s (i.e., the occupation number of species $i$) as
\begin{equation}
  g_i = w_i 2^{z_i},
  \label{eq:defgroup}
\end{equation}
where $w_i$ is the  group number  and $z_i$ the \textit{zoom number} (an integer equal to zero or larger) of species $i$. Thus, instead of tracking $g_i$ particles, we now simulate only $w_i$ groups, each containing $2^{z_i}$ particles. It is then the number of groups ($N_g$) that determines the collision rate per cycle. Because collisions are now between two groups of particles, involving $2^{z_i}$ and $2^{z_j}$ physical particles, coagulation is speeded up significantly. That is,
\begin{equation}
  N_g = \sum_i^{N_s} w_i; \qquad N_p = \sum_i^{N_s} w_i 2^{z_i},
\end{equation}
and we can, with a limited number of groups ($N_g$), represent large numbers of physical particles ($N_p$) by choosing appropriately high zoom factors $z_i$. This satisfies (I) and (II) and since each collision is now between two groups involving many particles we also avoid the inefficiency described in (III). Note that the base factor of 2 in \eq{defgroup} is arbitrary; however, it must be an integer ($g_i$ is integer) and 2 is probably most convenient. More critical are the choices for the zoom factors, $z_i$: these must be chosen to reflect the nature of the system that is under consideration (see \se{zoom}).

\begin{figure}
  \plotone{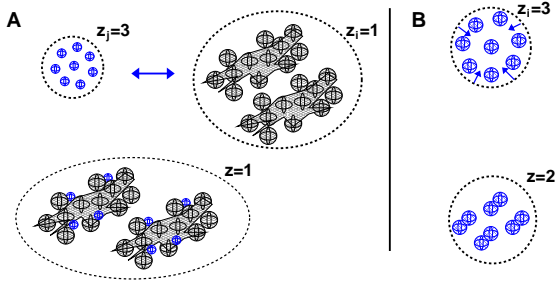}
  \caption{\label{fig:magnif}Illustration of collisions in the grouping algorithm. (\textit{A, top}) Two groups with $z_j = 3$ and $z_i = 1$, respectively, collide. The zoom factors ($z_i$) give the number of particles within the group ($2^{z_i}$). Less massive particles have higher zoom factors than particles of higher mass. All particles within the group are identical. (\textit{bottom}) Assuming sticking, the collision product is obtained by equal distribution of the collision partners. The end-product is a new species consisting of one group where the zoom factor corresponds to the lowest $z$ of the progenitor particles. \textit{(B)} Same, but for in-group collisions.}
\end{figure}
A group collision is illustrated in \fg{magnif}a. The first group is magnified three times ($z_j=3$), and the second has $z_i=1$ ($z_j\ge z_i$ in this and all other examples). The groups thus consist, respectively, of 8 and 2 identical particles. During the collision, assuming sticking, the particles are equally distributed over the groups so that we end up with one group consisting of two identical particles. The mass of the new particle then becomes $m_i+4m_j$ and the zoom factor $z=z_i=1$ (the lowest of the two; the grouping may always be adjusted if needed). Note that in \fg{magnif}a only particles of different groups collide: out-of group collisions. However, if their $\lambda_{ii}$ do not vanish, particles within a group can also collide: in-group collisions. This procedure is illustrated in \fg{magnif}b. The zoom number decreases by one in the final configuration (group collapse).

To generalize, for out-of-group collisions with $z_j\ge z_i$ every $i$-particle collides with $2^{z_j-z_i}$ $j$-particles. In total there are $2^{z_j}$ collisions, whereas for in-group collisions (\fg{magnif}b) there are $2^{z_i}/2$ collisions. Therefore, the collision rates of \eq{cr1} -- the rates corresponding to one collision -- are divided by $2^{z_j}$ and $2^{z_i-1}$, respectively, \ie
\begin{eqnarray}
  \nonumber
  \lambda_{ij}^G = g_i g_j C_{ij}/2^{z_j} = w_i w_j 2^{z_i} C_{ij}\qquad  (z_i \le z_j, i\neq j);\\
  \lambda_{ii}^G = \case{1}{2}g_i(g_i-1)C_{ii}/2^{z_i-1} = w_i(w_i 2^{z_i} -1) C_{ii}\qquad (i=j),
  \label{eq:cr2}
\end{eqnarray}
give the collision rates between two groups. Thus, although collisions between groups are less likely (the rate is decreased by a factor $2^{z_j}$), this is `compensated' by the $2^{z_j}$ particle-collisions that occur inside the group. 

However, grouping does introduce various sources of systematic errors:
\begin{enumerate}
\setlength{\itemsep}{-1mm}
\item the outcome of a collision between two groups is fixed; in reality, a distribution of configurations will be present;
\item it is assumed that the grouped collision rates obey Poisson statistics like the single particle collision rate;
\item the individual collision rates $\{ C_{ij} \}$ do not change \textit{during} the group collision process to ensure that \eq{cr2} is applicable.
\end{enumerate}
A fixed configuration for the outcome of the grouping process is required because it prevents (unwanted) proliferation of the diversity of species, which would increase the $N_s$ parameter and strain CPU requirements. However, the diversity of the simulation is already specified by the $N_s$ parameter and we may simply check how $N_s$ affects the accuracy of the collisional process. This point is similar to the second concern where all collisions are forced to occur at one point in time. Also, note that we must use the memoryless property of Poisson statistics; otherwise all collision rates must be adjusted at each timestep. However, it can be shown that, provided the collision rates stay constant, the grouped rates in \eq{cr2} will conserve the mean of the particle distribution if $i \ne j$ (see Appendix \ref{app:app1}). The last assumption is the weakest: the individual rates between the particles in principle change after each individual collision as the collision affects the properties of the particles. Therefore, expressions as $g_i g_j C_{ij}$ should actually be averaged over the number of collisions that make up the group process. This is of course a very tedious procedure with which we would go back to the one-by-one treatment of the particles. Rather, we prefer to introduce a new parameter, $f_\epsilon$, that limits the amount of mass accretion onto a particle during the group collisions, thereby strengthening the approximation that the $C_{ij}$ stay constant during the group collision process. 

Choosing the particle labels again such that $j$ has the high zoom factor ($z_j\ge z_i$) the fractional mass accretion on a single $i$ particle is $\Delta m_i/m_i = 2^{z_j-z_i}m_j/m_i$. From the preceding discussion it is this quantity that should be limited from below by a fixed value, $f_\epsilon$. One way to implement this is to reduce the $j$-particles that take part in the collision.  For example, in \fg{magnif}a this could mean that only 2 $j$-particles are involved, \ie\ only a fraction of the particles in a group. Let this fraction be denoted by $2^{-N_\epsilon}$ with $N_\epsilon\ge 0$ integer. Then, only $2^{z_j-z_i-N_\epsilon}$ $j$-particles are involved in the collision and the mass of an $i$ particle increases by $2^{z_j-z_i-N_\epsilon} m_j$, \ie
\begin{equation}
  \frac{\Delta m_i}{m_i} = \frac{2^{z_j-z_i-N_\epsilon}m_j}{m_i} \le f_\epsilon.
\end{equation}
Solving for the group splitting factor, $N_\epsilon$, gives
\begin{equation}
  N_\epsilon = \left[ -\log_2 (f_\epsilon 2^{z_i} m_i/ 2^{z_j}m_j) \right],
  \label{eq:Neps}
\end{equation}
where the square brackets denote truncation to integer numbers $\ge 0$. 
Consequently, the collision rates, instead of \eq{cr2}, now become
\begin{equation}
  \lambda_{ij}^G = g_i g_j C_{ij}/ 2^{z_j-N_\epsilon} = w_i w_j 2^{z_i+N_\epsilon} C_{ij};\qquad z_i \le z_j,
  \label{eq:cr3}
\end{equation}
(the collision rates for in-group collisions do not change since $N_\epsilon=0$). Another consequence is that the group numbers are real numbers, \ie\ since a fraction $2^{-N_\epsilon}$ of the $j$-group is removed, its group occupation number decreases accordingly, $w_j\rightarrow w_j-2^{-N_\epsilon}$. The $\{g_i\}$ cannot become fractional, though, since this would physically split a particle. Also, a situation where $0<w_i<1$ is not allowed, because $\Delta w = 1$ for the particle of the lowest zoom factor; when it arises we `demagnify' the group ($z_i\rightarrow z_i-1$) such that $w_i\ge 1$ (see below, \se{magdemag}).

\subsection{\label{sec:zoom}Choosing the zoom factors}
We have implemented the grouping method in such a way that the zoom factors, are free parameters, and there is indeed considerable freedom to choose (an algorithm for) the $\{z_i\}$. However, it is also an important choice as it will determine how \textit{efficient} the Monte Carlo simulation becomes. Here, with ``efficient'' it is meant that the number of species $N_s$ should not (suddenly) increase. A way that observes this constraint fairly well is to divide the mass inside the simulations equally over the groups. Additionally, we present a more general method, in which the zoom factors are determined by the particle distribution.

\subsubsection{Equal mass approach\label{sec:emm}}
\begin{figure*}
  \includegraphics[width=\textwidth]{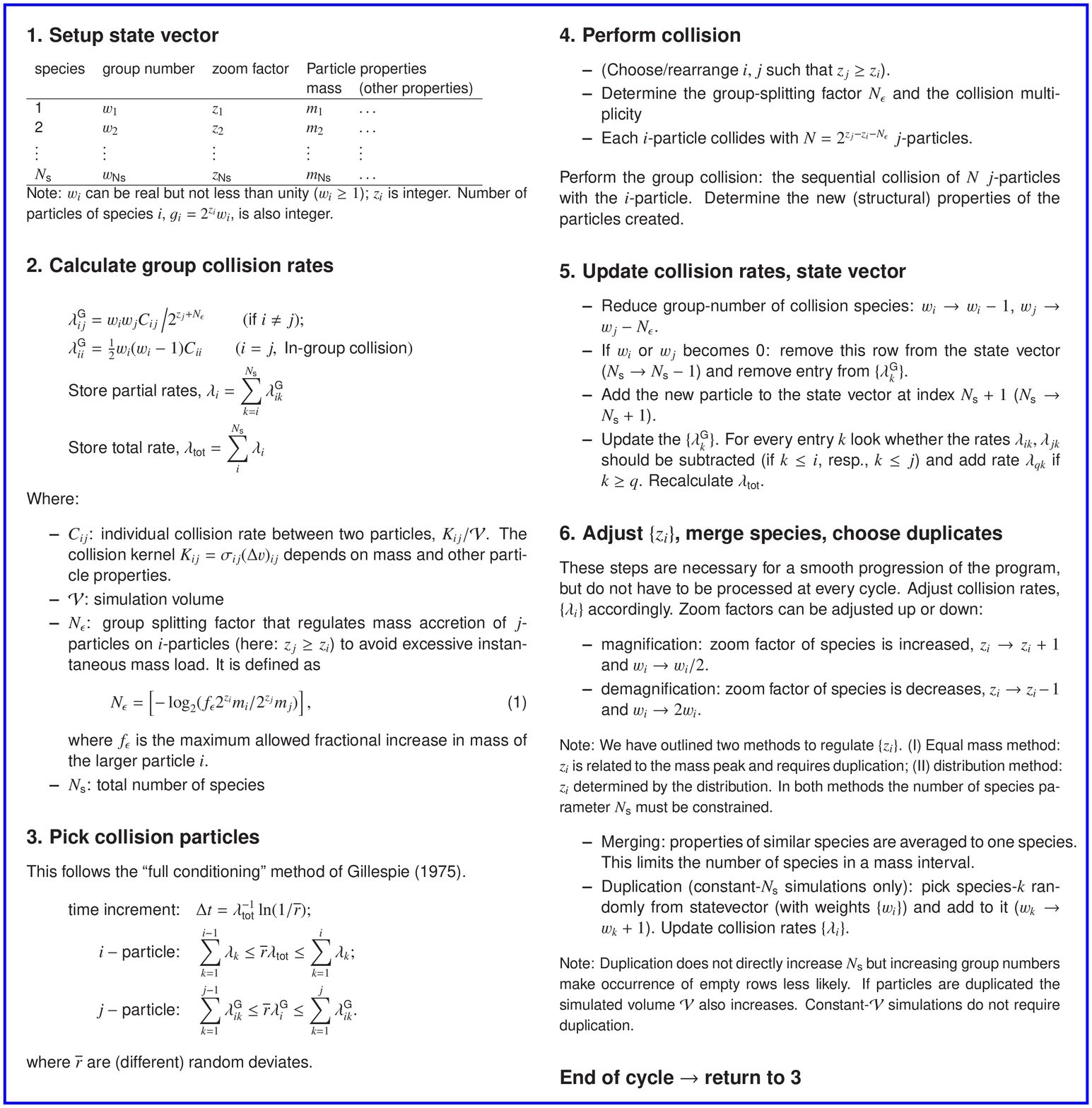}
  \caption{\label{fig:chart}Chart outline the steps required to implement the grouping algorithm.}
\end{figure*}
In this method, we assume there is a characteristic mass $m^\star$ after which the zoom factors are determined. Particles of mass larger than $m^\star$ are ungrouped ($z=0$), whereas particles of mass $m<m^\star$ have zoom factors such that the group mass is around $m^\star$, \ie
\begin{equation}
  z_i = \left[ \log_2 (m^\star/m_i) \right].
  \label{eq:zi}
\end{equation}
In the case of constant-$V$ simulations and monodisperse initial conditions of unity mass the simulation then starts with ${\cal N}/m^\star$ groups, where ${\cal N}$ is the initial number of particles. Therefore, ${\cal N}/m^\star$ is a measure for the number of species parameter $N_s$. The target value of $N_s$, $N_s^\ast$, and ${\cal N}$ then determine the value of $m^\star$.  For the constant-$N_s$ simulations, where the mass in the simulation varies, the mass peak $m_p$ is a natural value for the characteristic mass, $m^\star = m_p$.  It forces most of the groups to represent particles around the mass peak, with those of higher mass ungrouped. 

Thus, the equal mass approach resolves the mass density of the distribution, instead of the particle density as in conventional MC-methods. Once species becomes insignificant in terms of mass, $m_i2^{z_i}<m^\star$, occupancy number approaches unity ($w_i=1$) and the species will be ``removed'' from the simulation when it collides at the next collision, completely independent of the actual number of particles a group contains.

\subsubsection{Distribution approach\label{sec:dm}}
A different way to choose the zoom factors is to force the distribution to be resolved over its entire mass range. This can, for example, be done by dividing the distribution into exponentially-spaced mass bins and to choose the zoom factors such that there are a fixed number of groups per bin. Here the bins are exponentially distributed between a lower mass (most often the initial mass, $m_0$) and the largest mass $M_{L1}$ present in the simulation. For example, let $N_k$ be the total occupancy number of bin $k$, then the zoom numbers are determined by the requirement that an equal number of groups is present in each bin, \ie\  $N_k/2^{z_k}$ is constant for each bin $k$. There is no constraint on the number of species per bin, but as this quantity is always less than the number of groups it is limited as well. For a smoother progression of zoom factor with mass we allow for a linear interpolation within each bin, such that the zoom factors become a continuous, piecewise, and decreasing function of mass, $z(m)$.  Thus, $z$ is now determined, not by the particle or mass density, but by the sheer occurrence of particles of a certain mass.  The bins and zoom factors are dynamically adjusted, as the simulation progresses, and the distribution changes.

The distribution method has the consequence that (high-$m$) fluctuations are resolved very well, even if they are completely insignificant by number and mass. The zoom factors of these particles are much reduced in comparison with the equal mass method.  For example, if the $m_0=1$ particles are still at zoom levels of $z_0$ situations where $2^{z_0}m_0 \gg 2^{z_i}m_i$ are frequently encountered.  A collision between a massive and a $m=1$ group then takes place at a high value of $N_\epsilon$ (\eqp{Neps}) and collision rates involving high-$m$ groups increase (\eqp{cr3}). Since each collision creates a new species, the number of species ($N_s$) also increase rapidly -- too rapid, in fact -- and we are forced to merge species (\se{Ns}). Merging of species is therefore a necessary feature of the distribution method.

\subsection{(De)magnification\label{sec:magdemag}}
In the grouping method the zoom factors of the groups are determined in accordance with one of the two methods above. The zoom factors will therefore change during the simulation. When $z_i$ increases ($z_i\rightarrow z_i+1$) we speak of \textit{magnification}, whereas when $z_i$ decreases the group is \textit{demagnified} ($z_i\rightarrow z_i-1$). This dynamical adjustment of the zoom factors is the main reason why the method is capable of achieving very high dynamic range. Demagnification furthermore occurs when the occupancy number of a group becomes fractional, $0<w_i<1$. However, we note again that the actual number of particles of a species, $g_i$, is always integer.

There is a small price  to pay in terms of overhead for these re-adjustments. Because the $\{z_i\}$ are an argument of the collision rates (\eqp{cr3}) these rates (or, more accurately, the $\{\lambda_i\}$ quantities, as these are the ones that are stored in the computer's memory, $\sim N_s$ in total) have to be adjusted at each (de)magnification event. This may slow down the simulation by a factor of 2. Finally, we note that (de)magnification is, within the context of the grouping algorithm, an exact procedure; collision rates are not approximated and the total mass of the simulation stays conserved. 

\subsection{Grouping Summary\label{sec:groupsum}}
We have outlined an approach for a Monte Carlo coagulation scheme that can involve many particles -- allowing the growth to be orders of magnitude -- yet avoids the severe computational problems that would naturally arise with a single particle per collision treatment. In \fg{chart} we summarize the main steps of the algorithm, within the context of the grouping method. At its core lies the approximation that many identical particles are present, due to monodisperse initial conditions or the duplication mechanism, which can be efficiently grouped together or (in the distribution method) merged. Collisions are now between two groups of, possibly, many particles and this significantly speeds up the computation. While it is clear that such grouping inevitably approximates the collision process, we have advocated a fundamental shift in the implementation of MC methods, \ie\ to focus on the particles in a distribution that really matter -- \eg\ those around $m=m_p$ -- and we anticipate that the benefits resulting from the new emphasis will more than justify the approximations inherent to the grouping method. In any case, to validate the new method a comparison must be made with the analytic solutions of the classical kernels, after which the algorithm can be applied to more physically interesting kernels.

\section{Results}
\label{sec:results}
\begin{deluxetable*}{llllrrrr}
  \tablecaption{\label{tab:sims}List of simulations}
  \tablewidth{0pt}
  \tablehead{ \colhead{Kernel/description} & \colhead{Figure ref.}  & \colhead{Type}& \colhead{Grouping} &\colhead{$N_s^\ast/N_g^\ast$} &\colhead{$\cal{N}$} & \colhead{$f_\epsilon$} & \colhead{$N_\mathrm{sim}$} \\
              (1) & (2) & (3) & (4) & (5) & (6) & (7) & (8)}
    \startdata
    sum kernel              & \Fg{linnogroup}           & constant-$N_p$        & none          & $10^5$                & \nodata     & $10^{-4}$ & 40\\
    sum kernel              & \Fg{sum}a                 & constant-$N_s$        & equal mass    & $2\times10^4$         & \nodata     & $10^{-4}$ & 40\\
    sum kernel              & \Fg{sum}b                 & constant-${\cal V}$   & distribution  & $10^4$                & $10^{100}$  & $10^{-4}$ & 40\\
    product kernel          & \Fg{product}              & constant-${\cal V}$   & equal mass    & $2\times10^4$         & $10^{20}$   & $10^{-4}$ & 40\\
    product kernel          & \Fg{product2}             & constant-${\cal V}$   & distribution  & $10^3$                & $10^{20}$   & $10^{-4}$ & 40\\
    runaway kernels         & \Fg{runaway}              & constant-${\cal V}$   & distribution  & $\sim10^4-10^5$       & $300$--$10^{160}$ & $10^{-4}$ & 10-100\\
    sum: fragmentation      & \Fg{linearfrag}           & constant-$N_s$        & equal mass    & $2\times10^4$         & \nodata     & $10^{-3}$ & 1\tablenotemark{a}\\
    constant: fragmentation & \Fg{constantfrag}         & constant-$N_s$        & equal mass    & $10^4$                & \nodata     & $10^{-3}$ & 1\tablenotemark{a} 
    \enddata
    \tablenotetext{a}{In fragmentation simulations a steady state is reached in which the distribution is averaged $\sim 100$ times during the same simulation.}
    \tablecomments{Summary of all simulations. (1) Collision kernel. (2) Figure reference. (3) Simulation type: fixed or adjustable volume. (4) Grouping method: outlines in which way the zoom factors, $\{ z_i\}$, are chosen (no grouping, equal mass, or distribution method). (5) Target number of species (equal mass method), or groups (distribution method) in simulation. When multiple values apply the largest is given. (6) Initial number of monomers, or total mass in simulation for constant-${\cal V}$ simulations. (7) Fractional mass accretion. (8) Number of simulations performed. The initial distribution is monodisperse in all simulation with $m_0=1$. }
\end{deluxetable*}
In this section the grouping method -- both the equal mass approach, as well as the distribution method -- is tested in situations that require a high dynamic range.  Its performance is then discussed. First, the grouping method is tested against the sum kernel.  We show that both the equal-mass and the distribution methods are capable of following the analytic distribution, whereas conventional MC methods (without grouping) will fail. The grouping method is also tested against the product kernel, which is one of a class of kernels that produces runaway-growth behavior. Here, we will see that it is important to accurately follow the fluctuations of the mass distribution, which can only be done by the distribution method. More runaway-models are discussed in \se{runaway}. Finally, we also perform simulations including a simplified form of fragmentation, where the particles are catastrophically fragmented at the largest scale $m_\mathrm{frag}$ to be injected back at the smallest scale $m_0$ (\se{frag}). We test whether the grouping method can also be applied in these cases. An overview of the simulations is given in \Tb{sims}, including the adopted numerical parameters. These are chosen such that the CPU-time per simulation is typically several hours on a modern desktop machine ($1.8\ \mathrm{GHz}$ clock speed). Similar CPU-times apply for the choice of the numerical parameters of the simulations including grouping.

For the product and other runaway kernels, the system will after a certain time be dominated by a single particle. Hence, no self-similar solutions exist and in order to study the evolution properly, the simulation volume of these systems must be well defined. These simulations are therefore modeled using constant-volume simulations. In most cases the distribution method is used, which also requires a fixed simulation volume. In the other cases, we may assume that a small volume is representative of the parent distributions, which can then be modeled using constant-number simulations. The only exception is \fg{sum}b, where we have used the distribution method in order to compare it to the results of the equal-mass method simulations of \fg{sum}a. 

\subsection{\label{sec:sum}Test case I: the sum kernel}
\begin{figure}
  \plotone{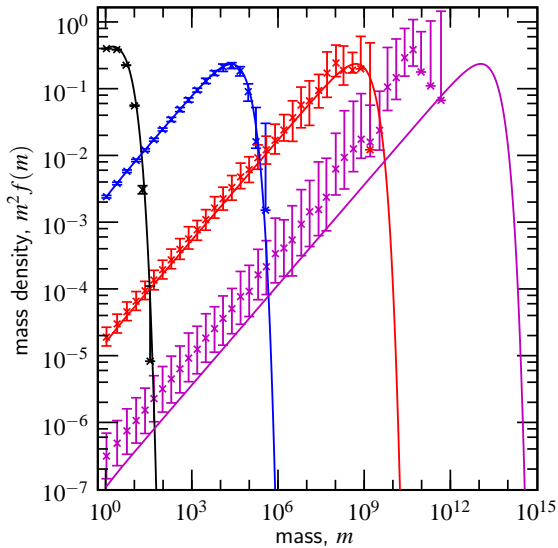}
  \caption{\label{fig:linnogroup}Results for the sum kernel, $K_{ij}=\case{1}{2}(m_i+m_j)$, \textit{without} grouping for $N_p=10^5$. Distributions are plotted at dimensionless times $t=1, 10, 20$ and $30$ and error bars show the spread over 40 simulations. Particles are binned at each factor of two in mass. The correspondence with the analytic results (solid lines) deteriorates over time. After $t\simeq 20$ the mass peak is no longer resolved by the simulation.}
\end{figure}
Before presenting the results of the grouping mechanism, a case without the grouping method is considered first, illustrating some of the `deficiencies' mentioned in \se{intro}. We will not consider the constant kernel, because in the constant kernel the dynamic range ($m_p/\langle m \rangle$) stays confined and a grouping method is not required. In \fg{linnogroup} the number of simulation particles, $N_p$, is kept constant at $10^5$. The distribution function is plotted at various times during its evolution, $t=1, 10, 20$ and $30$. Times are dimensionless, \ie\ in units of $(K_{00}n_0)^{-1}$ where $K_{00}$ and $n_0$ are the kernel and number density corresponding to the initially monodisperse conditions of particles of mass $m_0=1$. In \fg{linnogroup}, as well as in all other figures that show distributions from the MC-simulation, $f(m)$ is determined using exponentially-spaced bins every factor of 2 in mass. Each point shows the average and spread of in total 40 model runs. The analytic distribution functions at these times are given by the solid curves.

\begin{figure*}
  \plottwo{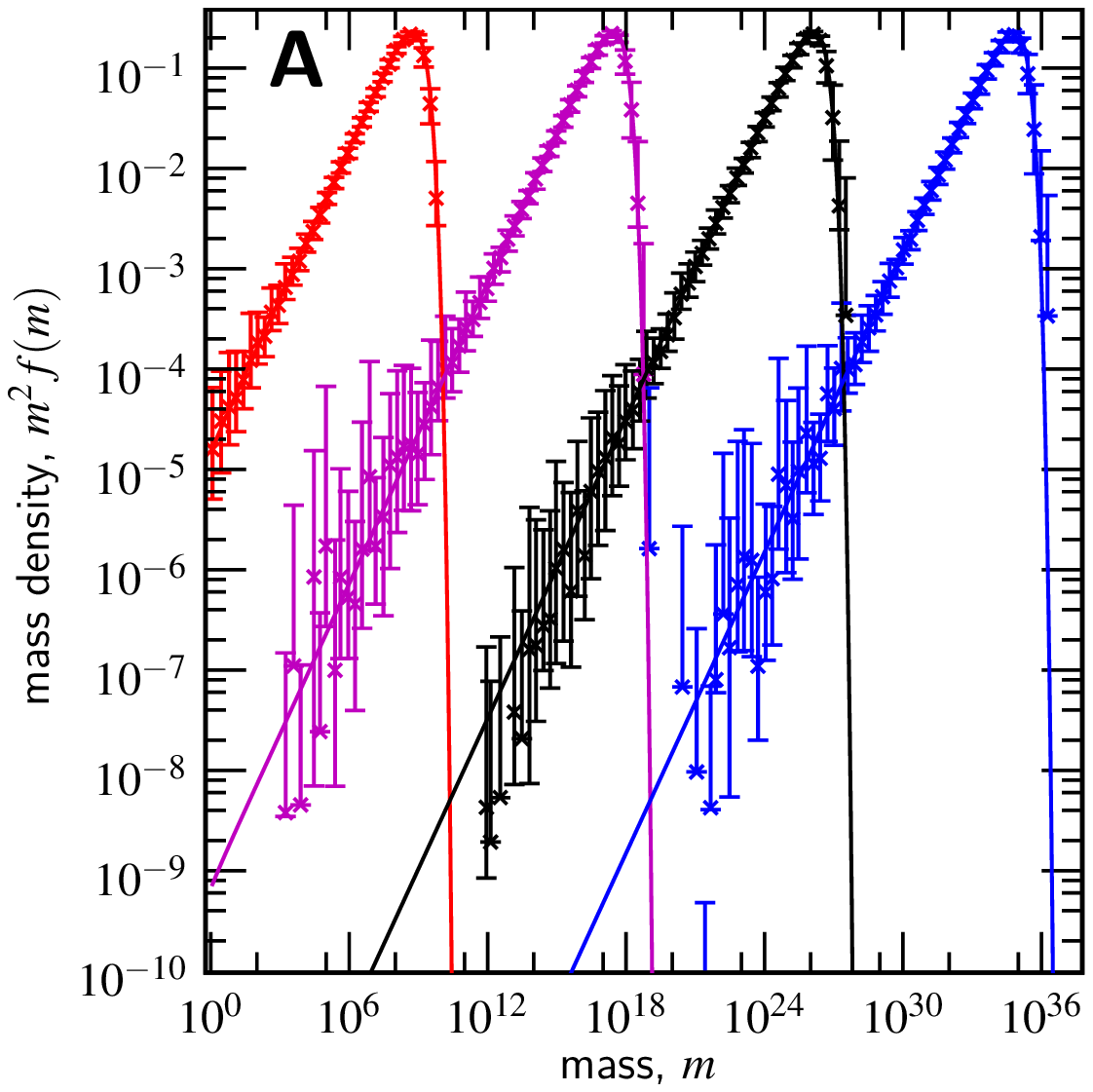}{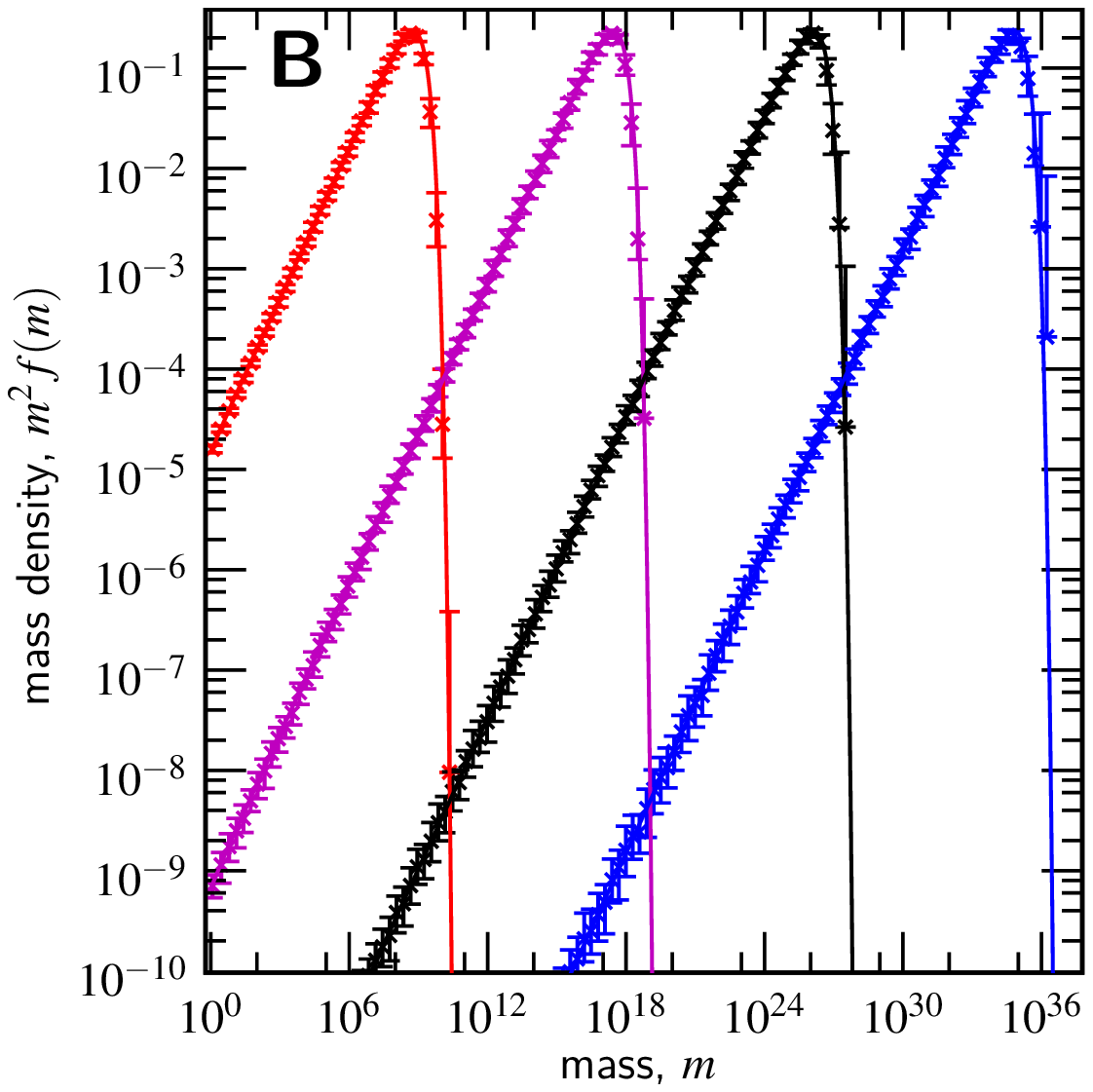}
  \caption{\label{fig:sum}Results for the sum kernel using the grouping algorithm. Particle distributions (symbols) are shown at times $t=20, 40, 60$ and $80$. Particles are binned at each factor of two in mass and results from 40 separate simulations are combined. The spread in the simulations is denoted by the error bars. \textit{(A)} Equal mass method, $N_s=20\,000$. Thanks to the grouping algorithm most of the groups are centered on the mass peak, which favours an accurate computation of the collisional evolution, without introducing systematic errors. \textit{(B)} distribution method with $N^\ast_g=10\,000$, distributed over 20 bins. The zoom factors are adjusted to resolve the distribution over its entire range. }
\end{figure*}
Initially, $t<10$, the agreement between the MC simulation and the analytic curve is very good. The mass range in the simulation is still rather modest and the $10^5$ particles suffice to represent all relevant masses. However, while the peak mass progresses steadily to higher $m$, the majority of particles stays behind at low masses; that is, although both $\langle m \rangle$ and $m_p$ evolve exponentially with time (\fg{distrib}), $m_p/\langle m \rangle$, which is a measure of the dynamic range, also increases exponentially. Since MC-methods sample the number density, increasingly fewer particles become associated with the mass peak, and the simulation looses numerical resolution. The effects of these low-number statistics are a bit alleviated by averaging over many different simulation runs but it is not merely a statistical deterioration: the mass peak dominates the collisional sweep-up of the system and the fluctuations caused by an inaccurate characterization blow up. There is especially a trend to lag the analytic solutions because, it is impossible to `recover' from a lagging distribution, because the duplication mechanism does not duplicate the particles that are not present. At $t=20$ these problems already become apparent when the spread is very high, especially around the mass peak, and there is a slight tendency to lag the analytic curve. At $t=30$ the effects are disastrous: $10^5$ particles are simply too few to sample a large enough volume to represent the large particles. 

Due to the duplication mechanism many duplicates are created and $N_s$ slowly decreases. This somewhat speeds up the computation at larger times since collisional rates scale with $N_s^2$. A more accurate result will be obtained by keeping $N_s$ constant \citep{2007A&A...461..215O}. However, now $N_p$ increases and the simulation suffers computational freeze-down as it progresses. Fundamentally, either approach -- the constant $N_p$ or constant $N_s$ -- is the same since each cycle only treats one collision and each collision is equally important. Clearly, this `socialist' approach is ineffective as \fg{linnogroup} illustrates: the more massive particles deserve more attention than their smaller brethren.

Results with the grouping method applied are shown in \fg{sum}. In \fg{sum}a we used the equal mass method for the groups (each group has a mass of $\sim m^\star = m_p$ if $m<m_p$ and particles with $m>m_p$ are not grouped) and choose the zoom factors accordingly (see \eqp{zi}). 
The number of species, $N_s$, is stabilized at a target number of $N_s^\ast=2\times10^4$ through the duplication mechanism. That is, if $N_s$ due to coagulation falls below $N_s^\ast$, duplicates are added to the system until $N_s\ge N_s^*$. The simulation total mass and volume then increase, but their ratio (the mass density) is fixed. As $m_p$ becomes larger during the simulation, zoom factors of particles that do not take part in collisions will increase, according to \eq{zi}. This `magnification' causes the occupancy numbers $w_i$ to stay low, despite the ongoing duplication events. When $w_i$ reaches 1, the species will disappear at the next collision, freeing up this slot for more massive particles. The grouping thus forces the particles to act collectively. This procedure prevents the systematic errors ($N_p$ constant, \fg{linnogroup}) or computational freeze down (constant $N_s$) of the `1 collision per cycle' approach of the conventional MC-methods.

\Fg{sum}a shows that the grouping mechanism resolves the mass peak well during its entire simulation. 
The agreement between the analytic and numerical method remains very good during the entire simulation. \Fg{sum}a proves that once the mass peak is resolved, the simulation will accurately compute the collisional evolution, and systematic errors as in \fg{linnogroup} are prevented. Thus, the resolution of the mass peak is key. However, at lower mass densities the groups disappear from the simulation as they do not have sufficient mass. 

These low-$m$ statistics can be accurately followed when we switch the method for assigning the group's zoom factor $\{z_i\}$ to the distribution method, in which the zoom factors are adjusted such that there are an equal number of groups per exponential mass interval. The zoom number for the low-$m$ particles is then much lower in comparison with the equal mass method. \Fg{sum}b shows that the mass density as well as number density are accurately followed over the entire distribution. It can be seen in \fg{sum}b that the match to the high-$m$ tail slightly deteriorates over time. This is a consequence of the equal spacing in log space of the bins that determine the zoom factors. The high-$m$ tail, which falls off exponentially, becomes associated with only a single bin, and is then less well resolved. A more flexible placement of the bins is expected to give an even better correspondance.  
Note again that the bins are only needed to determine the zoom factors; there's no actual binning of the particles in the MC-simulation.

Although \fg{sum}b more precisely follows the distribution -- \ie\ it both follows the mass density as well as the number density -- it requires to \textit{merge} species in order to restrict their numbers (see \se{Ns}).  For these test cases, in which only mass is a parameter, \fg{sum}b shows that this is not problematic. However, when structural parameters are involved, the merging procedure is less obvious; \ie\ are species of comparable mass merged or, say, species of comparable porosity? 
The merging principle is, therefore, somewhat against the spirit of MC-simulations, as there is no physical equivalent of 'merging' in nature. Thus, for a MC simulation with many structural parameters, where it is not required to resolve the number density, \fg{sum}a is probably the preferred method. In the other test cases of this section we will therefore first pursue whether these can be modeled by the equal-mass method.  

\subsection{\label{sec:product}Test case II: the product kernel}
The kernel $K_{ij} = m_im_j$ is the only runaway kernel with an analytic solution. In models that show runaway-behavior a particle will appear that separates from the continuous distribution to consume, in a relatively short time, all the available mass. This separation is an intrinsic feature of the kernel and is not caused by a poor sampling of the distribution. The point at which this separation occurs, $t_R$, depends on the properties of the kernel and, in some cases, on the initial number of particles (${\cal N}$). In the case of the product kernel it always occurs at $t=1$, independent of ${\cal N}$. The runaway growth properties of other kernels are discussed in \se{runaway}.

In \se{intro} it was noted that the Smoluchowski coagulation equation becomes invalid in the case of the runaway kernel: it keeps assuming that the distribution is continuous and it is therefore ill-suited to describe a discrete system caused by the runaway particle. \citet{1990Icar...88..336W} has adjusted the original (incorrect) solution of \citet{1971Trubnikov} by inclusion of a term for the runaway body in \eq{smoluchowski}. Although not mathematically rigorous, this approach captures the physics of the runaway: as soon as the (incorrect) numerical solution exhibits a $f(m)\propto m^{-5/2}$ power-law tail, the interaction between the higher mass bins causes the runaway growth. In the Monte Carlo simulations we also expect that the appearance of a power-law tail is a signature that runaway growth is incipient.

\begin{figure}
  \plotone{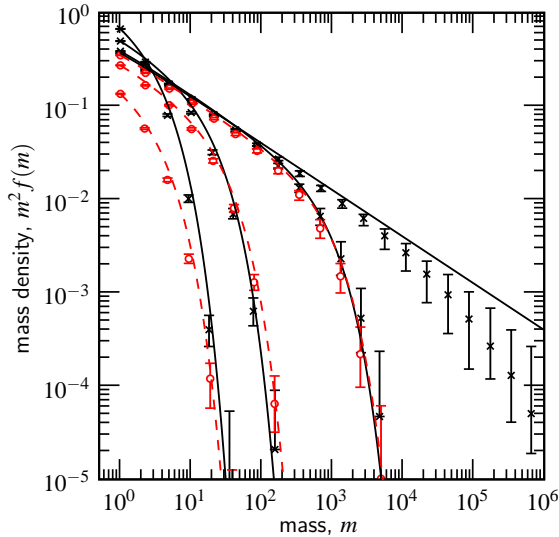}
  \caption{\label{fig:product}Results for the product kernel, $K_{ij}=m_im_j$, with ${\cal N} = 10^{20}$. The simulated distribution is plotted at times $t=0.4$, 0.7, 0.95, 1.0 (from left to right, \textit{crosses}) and $t=1.05$, 1.3 and 2.0 (from right to left, \textit{open circles}). The analytic solution is plotted by solid curves for $t\le1$ and by the dashed curves for $t>1$ (the $t=0.95$ and $t=1.05$ curves almost overlap). Error bars denote the spread in the 40 simulations. After $t\approx 1$ a runaway particle of mass $m\sim {\cal N}$ appears that consumes most of the mass in the system.}
\end{figure}
In \fg{product} the distributions corresponding to the product kernel with ${\cal N}=10^{20}$ are plotted at dimensionless times $t = 0.4, 0.7, 0.95, 1.0, 1.05, 1.3$ and $2.0$, using the equal mass method. Since we start the simulation already with a huge number of particles, large zoom numbers are required. We have fixed the mass peak at $m_\star = {\cal N}/10^4$, which causes the particles to be sufficiently magnified, except for the runaway particle. (The $10^4$ factor may be thought of as a fudge parameter akin to $N_s$ or $N_p$ in the constant-$N$ simulations.) Theoretical curves correspond to the \citet{1971Trubnikov} solutions which, as \citet{1990Icar...88..336W} showed, are still valid for the continuous part of the distribution. At $t=1$, runaway growth is expected and the continuous power-law distribution reaches its furthest extent: the $-5/2$ power-law exponent. After the runaway particle has separated the continuous distribution shrinks, especially in the $-x$ direction since the more massive particles are consumed rapidly by the runaway body, which itself is not shown in \fg{product}.

\begin{figure}
  \plotone{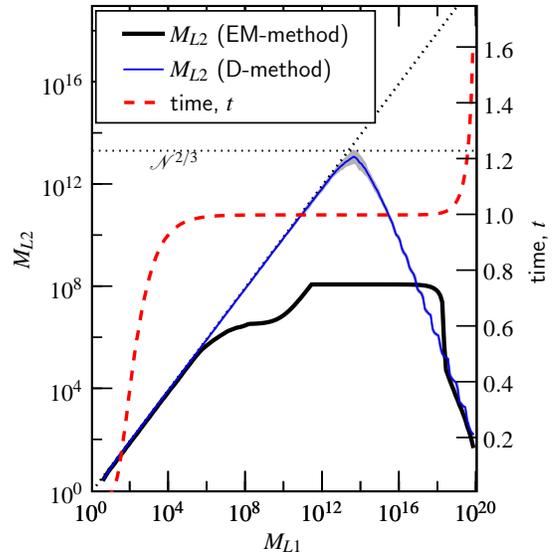}
  \caption{\label{fig:product2} Statistics for the simulations involving the product kernel. As function of the mass of the largest species, $M_{L1}$, are shown: \sumi\ the mass of the second-largest species, $M_{L2}$, according to the equal-mass method (\textit{thick solid curve}) and the distribution method (\textit{thin solid curve}); and, \sumii\ the time (\textit{dashed curve}). Values are averaged over 40 runs. The auxiliary lines $M_{L2}=M_{L1}$ and $M_{L2}={\cal N}^{2/3}$ are also given (\textit{dotted lines}). 
}
\end{figure}
The runaway growth is generally well modeled by the Monte Carlo method. The distribution becomes most sensitive near $t=1$ where a small change in time means a big change in the high-$m$ distribution function. Due to the stochastic behavior of the Monte Carlo method, the precise timing of the runaway event does not exactly coincide with $t=1$. Therefore, the $t=1$ distribution lags the theoretical curve since at $t=1$ either runaway growth did not yet occur (and the distribution still has to reach the power-law curve) or has already happened (and the distribution is on its way back from the power-law curve). In \fg{product2} the relation between the two most massive groups is denoted by the thick solid line, together with the simulation time $t$ as function of $M_{L1}$. \Fg{product2} shows that until $t=1$ there is little difference between the mass of the most massive particle -- actually group -- $M_{L1}$ and the second-most massive group $M_{L2}$. However, as runaway is reached at $M_{L1}\sim 10^5$ the masses diverge very quickly. 

These results, however, are in disagreement with the theoretical study of \citet{1994Icar..107..404T}. Comparing the solutions to the stochastic and the statistical equation they find that, in the case of the product kernel, the statistical equation (\eqp{smoluchowski}) becomes invalid after $M_{L1} \sim {\cal N}^{2/3}$. It is only at this point that $M_{L1}$ and $M_{L2}$ should start to diverge. However, in the equal-mass method (\fg{product2}, \textit{thick line}) the divergence happens much earlier: $M_{L2}$ never reaches beyond $10^8 \ll (10^{20})^{2/3}$. This discrepancy can be explained as an artifact of the equal mass method. Because each group is still of considerable mass, $m^\star \sim 10^{16}$, runaway growth is just the collapse of one group by a continuous series of in-group collisions (\fg{magnif}b); that is, until $z_i=0$, the $M_{L1}$ group does not correspond to a single particle and therefore never really separates. It is thus not a proper modeling of a runaway process. In order to accurately follow the behavior of the particles during runaway growth, the groups should be demagnified much more rapidly than \eq{zi} allows. It is perhaps surprising that the equal-mass method still works fine -- \ie\ its distribution agrees with the analytical case (\fg{product}) -- despite the fact that it does not resolve the runaway particles. However, in general this is not the case: the fluctuations present at high-$m$ require more resolution than what is attributed to them through the equal representation of mass density (\eqp{zi}). 

In \fg{product2} the $M_{L1}, M_{L2}$ statistics of the product kernel are recalculated with the distribution method for the zoom factors (thin solid line). According to the \citet{1994Icar..107..404T} prediction $M_{L1}$ and $M_{L2}$ should then separate at $M_{L1} \sim {\cal N}^{2/3}$. This is exactly the trend seen in \fg{product2}. We therefore conclude that the distribution method is the appropriate choice to compute the collisional evolution of runaway kernels.

\subsection{\label{sec:runaway}Strong Runaway kernels}
\begin{deluxetable}{lll}
  \tablewidth{0pt}
  \tablecaption{\label{tab:modes}Runaway models}
  \tablehead{ \colhead{type} & \colhead{conditions} & \colhead{growth mode} }
  \startdata
    \sumi   & $\nu\le1, \beta\le 1$     & orderly growth \\
    \sumii  & $\nu\le1, \beta>1$        & runaway growth at $t_0$ \\
    \sumiii & $\nu>1$                     & instantaneous runaway growth\tablenotemark{a}
  \enddata
  \tablenotetext{a}{Instantaneous in the limit ${\cal N}\rightarrow \infty$.}
  \tablecomments{Properties of runaway models according to the study of \citet{2000Icar..143...74L}. The exponents $\nu$ and $\mu$ give the mass dependence of the kernel on a heavy and light particle, respectively, with $\beta = \nu + \mu$ the total mass dependence, $K_ij \propto m_i^\nu m_j^\mu\ (m_i \gg m_j)$.}
\end{deluxetable}
There is an extensive literature on the phenomenon of runaway growth, especially mathematical (${\cal N}\rightarrow \infty$), where the phenomenon is termed \textit{gelation}. In general, three growth modes can be considered (see \Tb{modes}): \sumi\ kernels where runaway growth does not appear, as in the constant or sum kernel; \sumii\ kernels that exhibit growth after some characteristic time, as in the product kernel; and \sumiii\ kernels in which gelation is \textit{instantaneous}\footnote{We note that in a physical situation where runaway growth tends towards instantaneity, causality should not be violated. As we will see, ${\cal N}$, even in astrophysical circumstances, does not become so large that the process is instantaneous.}. The occurrence of instantaneous gelation for the class of kernels $K\propto (m_i m_j)^\nu$ with $\nu>1$ was conjectured by \citet{spouge1985mcr} on the basis of MC-coagulation experiments and was proved by \citet{1998CMaPh.194..541J}. \citet{2000Icar..143...74L} illustrates these distinct modes of solution by a simple coagulation model, in which the distribution is approximated by a characteristic mass $m$, \eg\ the peak mass, and a test particle of mass $M>m$. Particles of mass $m$ dominate the total mass and hence $n(m)\propto m^{-1}$. Furthermore, the coagulation kernel is characterized by two exponents, $\nu$ and $\mu$, that give the dependence of $K(m_i, m_j)$ in the case of a heavy ($i$) and a light particle ($j$), \ie\ $K_{ij} \propto m_i^\nu m_j^\mu$, for $m_i \gg m_j$. The total mass dependence is then given by $\beta = \mu+\nu$. For example, in the case of the product kernel $\nu=\mu=1$ and $\beta=2$. Setting the physical density equal to unity, the equations for the simplified merger problem become \citep{2001JPhA...3410219H}
\begin{eqnarray}
  \label{eq:dmdt}
    \frac{\mathrm{d}m}{\mathrm{d}t} = \case{1}{2} m^\beta;\\
    \label{eq:dMdt}
    \frac{\mathrm{d}M}{\mathrm{d}t} = m^\mu M^\nu.
\end{eqnarray}
  These can also be combined to obtain
\begin{equation}
  \frac{\mathrm{d}M}{\mathrm{d}m} = 2 \left( \frac{M}{m} \right)^\nu.
  \label{eq:dMdm}
\end{equation}
\Eq{dmdt} can be solved as
\begin{equation}
  m = \cases{ 
        \left[ 1 + \case{1}{2}(1-\beta) \right]^{1/(1-\beta)} & for $\beta\neq 1$; \cr
        \exp[\case{1}{2}t]                                         & for $\beta=1$,}
    \label{eq:mt}
\end{equation}
where we have for simplicity taken $m_0=m(t=0)=1$. \Eq{dMdm} can be solved in terms of $m$ as
\begin{equation}
  M(m) = \cases{
    \left( M_0^{1-\nu} + 2m^{1-\nu} -2 \right)^{1/(1-\nu)} & for $\nu\neq 1$; \cr
    M_0 m^2                                                & for $\nu=1$,}
    \label{eq:Mt}
\end{equation}
where $M_0>1$ is the inital mass of the test particle. Three qualitatively distinct growth modes follow from these equations: \sumi\ if $\nu\le1$ and $\beta\le1$ both \eq{mt} and \eq{Mt} have finite solutions at all $t$: orderly growth. \sumii\ If $\nu\le1$ and $\beta>1$, \eq{mt} reaches a singularity at a time $t=t_R=2/(\beta-1)>0$,\footnote{Note that this expression disagrees with the product kernel in which $\beta=2$: the simplified merger model does not give the correct prefactor, see \citet{2001JPhA...3410219H}.} while \eq{Mt} does not: both $m$ and $M$ become infinite at the same time $t_R$ that does not depend on $M_0$. \sumiii\ If $\nu>1$, \eq{Mt} reaches infinity for a finite value of $m$. Furthermore, the gelation time \textit{decreases} with increasing $M_0$, $t_R \simeq M_0^{1-\nu}/(\nu-1)$. Then, if the initial mass in the system goes to infinity, a sufficiently strong fluctuation (large $M_0$) will always be present to satisfy the singularity condition for $M(t)$ even as $t\rightarrow0$; this is the physical reason behind the instantaneous gelation conjecture.

Here, we will concentrate on the case $\nu>1$ and test the instantaneous gelation conjecture for kernels $K=(m_im_j)^\nu$ with $\nu=2,3$. \citet{2001Icar..150..314M} extend the simplified merger model discussed above to a model with only two kinds of particles -- one predator and many preys -- and one collisional process (the predator feeds of preys). This is the monotrophic model, \eg\ one particle (`mono') that nourishes (`trophein') of the others. Considering this model \citet{2001Icar..150..314M} compute the probability that the predator occupies any finite mass, which decreases with increasing time. Consequently, the gelation time $t_R$ is computed as
\begin{equation}
  t_R \propto \left( \log {\cal N} \right)^{1-\nu},
  \label{eq:tR}
\end{equation}
where ${\cal N}$ is the initial number of particles present. Thus, the timescale for runaway growth decreases as function of the initial number of particles in the system, but only logarithmically. 
\Eq{tR} may be anticipated from the discussion above on the simple merger model. Consider a distribution of particle masses $N(m)$ at a fixed time. It is quite likely to expect that the tail of the distribution is exponentially suppressed as in the case with Poisson statistics, \ie\ $N(m)\propto {\cal N} \exp[-m]$ with ${\cal N}$ the total number of particles. Setting $N=1$ we thus obtain $M_0 \sim n\sim \log{\cal N}$. Hence, the logarithmic dependence on the number of particles.
\citet{2001Icar..150..314M} test this model by Monte Carlo simulations and found qualitative agreement: the gelation time decreases as function of $\log {\cal N}$, although the power-law exponent is not precisely that of \eq{tR}: $1-\nu$. However, the simulations of \citet{2001Icar..150..314M} were limited in their dynamic range; we will extend their study to much larger values of ${\cal N}$ to see whether \eq{tR} holds over a sizeable range of $\log {\cal N}$.

Note that as conjectured earlier, a good census of the fluctuations is the key to an accurate modeling of the collisional evolution. Therefore, the grouping must then be implemented by the distribution method. There is, however, another catch. At $t=0$ the number of groups of the monodisperse species is $w_0 \sim N_g^\ast \ll {\cal N}$ and the collision rate for a collision between two monomers according to \eq{cr2} is $\lambda_{11} \sim  {\cal N} w_0 C_{11} \sim w_0$. Then, the timescale for collisions involving monomers is $t_{11}=\lambda_{11}^{-1} \sim w_0^{-1}$, which can, however, become comparable to, or larger than, the runaway timescale of \eq{tR}. Therefore, the monodisperse groups have to be demagnified such that $t_{11}\ll t_R$. By forcing the appropriate collision timescale for the monodisperse collisions, the fluctuations are adequately resolved, but it also means that the CPU-time increases, especially at large ${\cal N}$. For the $\nu=2$ simulations we used a number of $N_g^\ast = 60\,000$, whereas in the $\nu=3$ simulations we take $N_g^\ast = 300\,000$. At $\nu=3$, runaway occurs at a lower value of $M_{L1}$ and we are allowed to start with a higher number of groups. However, the discrepancy between the simulations at different values of ${\cal N}$ becomes large in terms of CPU-time: the highest-${\cal N}$ simulations take about half a day to finish.

\begin{figure}
  \plotone{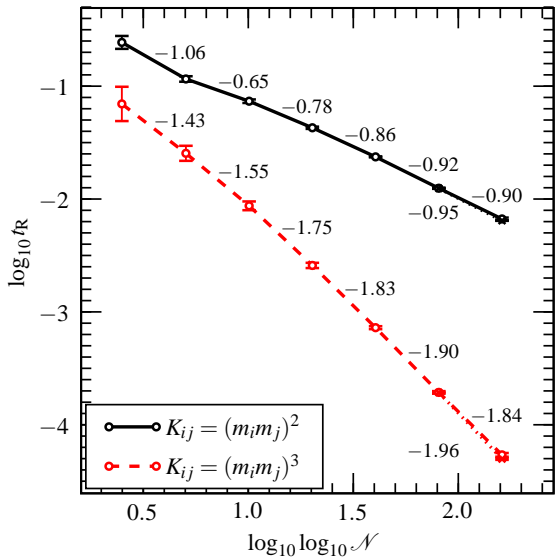}
  \caption{\label{fig:runaway}The runaway time $t_R$ as function of the initial number of particles ${\cal N}$ for the kernel $K_{ij} = (m_i m_j)^\nu$ with $\nu=2$ (solid line) and $\nu=3$ (dashed line). Simulations are run at ${\cal N} = 300, 10^5, 10^{10}, 10^{20}, 10^{80}$ and $10^{160}$. The power-law index is indicated at every interval of $\log {\cal N}$. The agreement with the theoretically expected value for the power-law index, $1-\nu$, becomes better at higher ${\cal N}$. At ${\cal N} = 10^{160}$ a simulation at a higher numerical resolution gives an even steeper slope (indicated below the line) confirming the converging trend towards, respectively, $-1$ and $-2$.}
\end{figure}
The results for the kernels $K=(m_i m_j)^2$ and $K=(m_i m_j)^3$ are presented in \fg{runaway}. We extend the model of \citet{2001Icar..150..314M} to an unprecedented $10^{160}$ number of particles (about the square of the number of particles in the universe). To speed up the simulations at high ${\cal N}$ we have decreased the mass-ratio parameter $f_\epsilon$ to $10^{-3}$ (which we checked to be still sufficiently accurate). The number of simulations varies between $N_\mathrm{sim}=100$ at low ${\cal N}$ and $N_\mathrm{sim}=10$ at ${\cal N}=10^{160}$. We stop the simulations when the ratio between the two largest masses becomes $M_{L1}/M_{L2}\ge10^4$, which is a measure of $t_R$. By this time the simulation has either already finished or has reached it asymptotic limit. 

According to the analytic prediction \citep[\eqp{tR},][]{2001Icar..150..314M}  the $t_R$-$\log {\cal N}$ power-law is $-1$ and $-2$ for $\nu=2$ and $\nu=3$, respectively. In \fg{runaway} the piecewise exponent is indicated (number above the lines), which shows a declining trend, except for the ${\cal N}=10^{160}$ point. To check convergence, we have additionally performed five simulations at a higher numerical resolution for ${\cal N}=10^{80}$ and ${\cal N}=10^{160}$ ($N_g^\ast = 2\times10^5$ for the $\nu=2$ simulations and $N_g^\ast = 2\times10^6$ for the $\nu=3$ simulations). For ${\cal N}=10^{80}$ it was found that the results had converged, but at ${\cal N} = 10^{160}$ a reduced $t_R$ compared to the lower resolution simulations was found. This increased the local slope between ${\cal N}=10^{80}$ and ${\cal N}=10^{160}$ (dotted lines, hardly distinguishable from the dashed lines, with the slope indicated below the line). Altogether, \fg{runaway} suggests that the agreement between the numerical simulation and the monotrophic model becomes better as ${\cal N}$ increases, a conjecture already made by \citet{2001Icar..150..314M}.

\subsection{\label{sec:frag}Fragmentation}
The grouping method can also be used in steady-state situations where fragmentation balances coagulation. From the simulations in \se{sum} we expect that for the distribution method this is not a problem, as it follows both the number as well as the mass density. We will therefore concentrate our efforts on the equal-mass method. Fragmentation events in Monte Carlo may seem problematic, because catastrophic destruction can result in the creation of many small particles that consume computational resources. However, due to the grouping method we now have a natural way to deal with fragmentation. In this section we consider a test case of complete destruction -- meaning: the breakup of a particle into its smallest constituents (monomers) -- that occurs once a mass threshold, $m_\mathrm{frag}$, is exceeded. We then solve for the emergent, steady-state, mass distribution. Note that due to stochastic fluctuations the number distribution is never really steady-state. We will therefore average over many distributions during the same simulation run to obtain the `steady-state' mass spectrum.

\begin{figure*}
  \plottwo{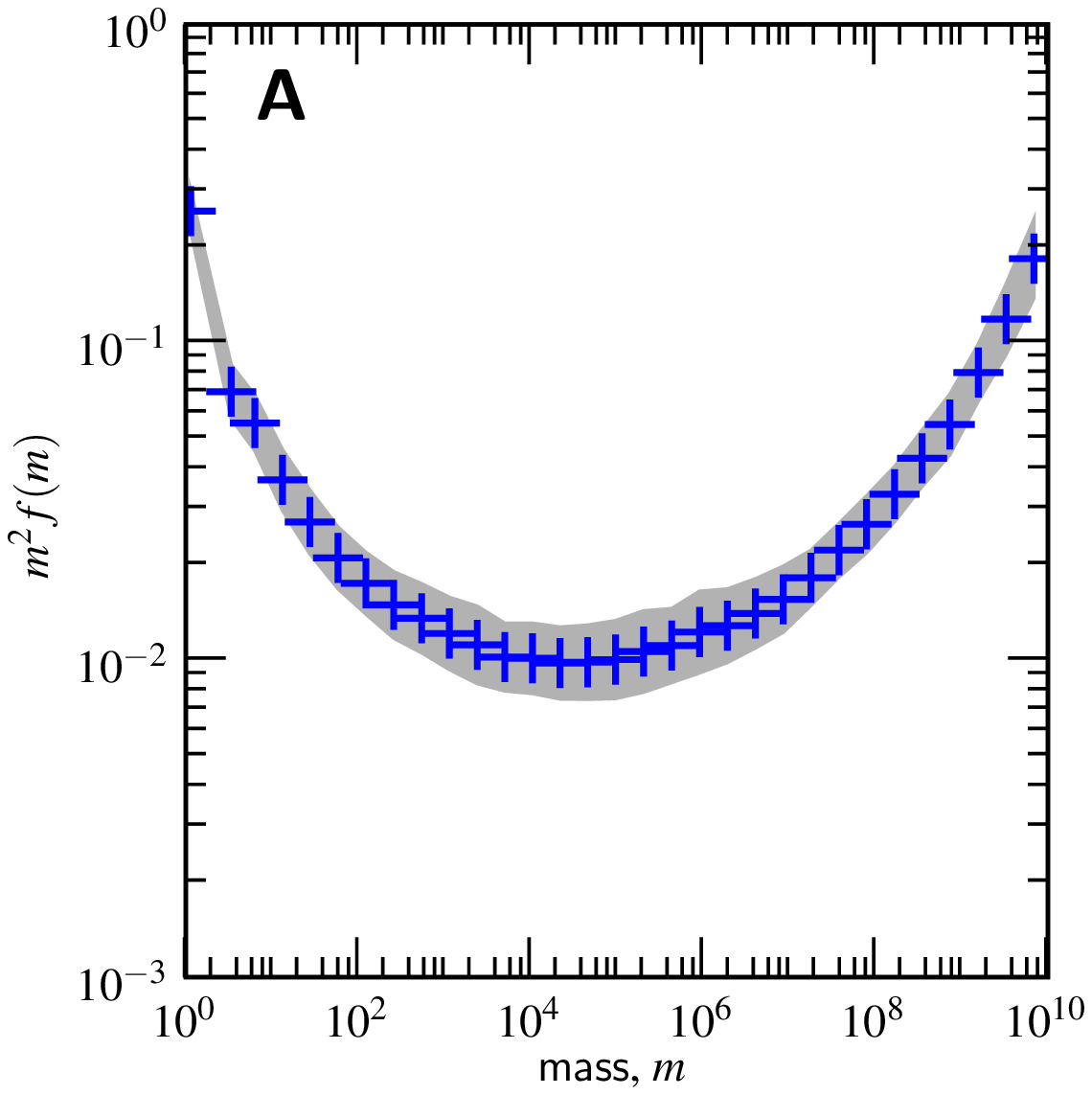}{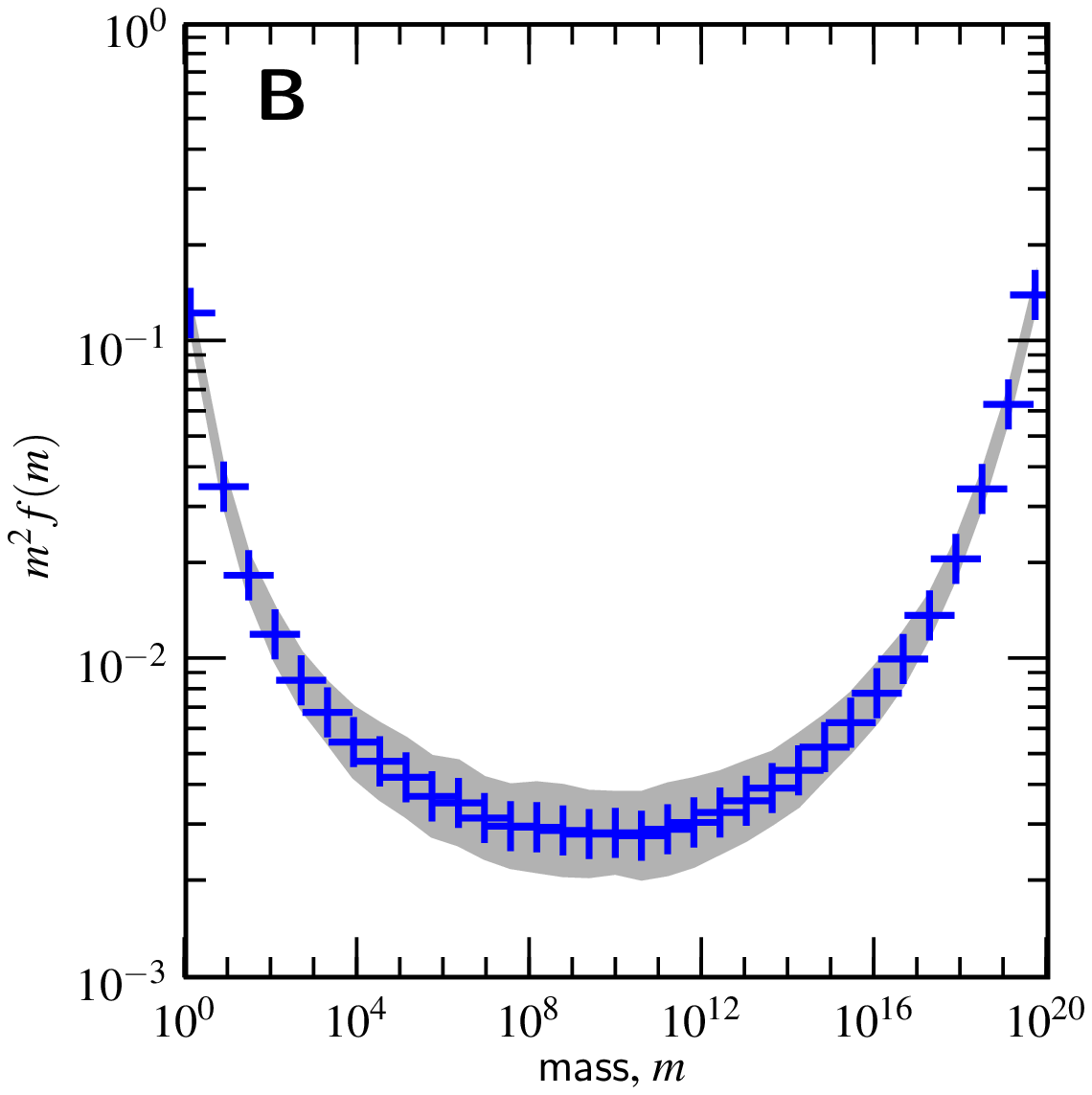}
  \caption{\label{fig:linearfrag}Steady state distributions of the sum kernel. At $m=m_\mathrm{frag}$ particles are catastrophically disrupted into monomers of mass $m=m_0=1$. $(A)$ $m_\mathrm{frag}=10^{10}$ and $N_s^\ast=10^4$; \textit{(B)} $m_\mathrm{frag}=10^{20}$ and $N_s^\ast=2\times10^4$. Distributions are averaged over various times during the steady-state process. Crosses denote this average over $\sim 100$ distributions. The spread is indicated by the grey shading.}
\end{figure*}

In \fg{linearfrag} the results of the sum kernel, $K_{ij} = m_i + m_j$, are given. In \fg{linearfrag}a the fragmentation threshold is $m_\mathrm{frag}=10^{10}$. The thick crosses denote the average over $\sim 100$ distributions during the simulation after the initial distribution has relaxed to semi-steady state. The volume of the simulation is stabilized when it incorporates $N_s^\ast=10^4$ species at ${\cal V} = 2\times10^{12} \gg m_\mathrm{frag}$. \Fg{linearfrag}b presents the result for an even larger value of $m_\mathrm{frag}=10^{20}$. The same characteristic distribution -- resembling a bath tub -- is found. These results suggest that, for the sum kernel with catastrophic fragmentation at a cut-off mass, most of the mass density becomes located near the end points. The system may then be approximated by two states at $m = m_0 \sim1$ and $m=m_\mathrm{f}\sim m_\mathrm{frag}$. The mass-loss of the monomer state due to collisions with the high-$m$ state is $m \mathrm{d}m/\mathrm{d}t|^- = n_0 n_\mathrm{f} m_\mathrm{f}$ and the gain due to fragmentation within the high mass bin becomes $m \mathrm{d}m/\mathrm{d}t|^+ = n_\mathrm{f}^2 m_f^2$. Balancing these we get $n_0 = n_\mathrm{f} m_\mathrm{f}$, \ie\ equal mass densities for the monomer and high-$m$ state.

However, with other kernels the mass distribution due to fragmentation may also emerge as a power-law. To see this we adopt a simple model \citep{PhpsRevE..63..046112} in which the mass sweep-up, $\dot{m}$, of a particle of mass $m$ due to particles of lower mass $m'<m$ is assumed to change only the mass in a continuous way. That is,
\begin{equation}
  \dot{m} = \int_{m_0}^m dm' K(m,m') m' f(m'),
  \label{eq:mdot}
\end{equation}
will shift particles of mass $m$ by an amount $\Delta m = \dot{m}\Delta t$. The total particle change per unit time in the interval $[m, m+\Delta m]$ becomes $-\Delta[\dot{m}f(m)]$ due to gradients in $\dot{m}$ and the distribution function. The other side of the distribution then acts as a sink to the particles of mass $m$, \ie\ $R(m) f(m) \Delta m$ is the number of particles in $[m,m+\Delta m]$ that is `consumed' by the more massive particles, with $R(m)$
\begin{equation}
  R(m) = \int_m^{m_\mathrm{frag}} dm' K(m,m') f(m'),
  \label{eq:Remm}
\end{equation}
the removal rate. For a steady state we must therefore have
\begin{equation}
  \frac{\partial (\dot{m} f(m))}{\partial m} = - f(m) R(m).
  \label{eq:ssfrag}
\end{equation}
Assuming a power-law for $f$, $f\propto m^{\alpha}$, \eq{ssfrag} can be solved for $\alpha$ by extending the limits of the integrals in \eq{mdot} and \eq{Remm} to 0 and $\infty$, respectively \citep{PhpsRevE..63..046112}. If $|\beta|<1$ the integrals converge and the resulting power-law exponent is $\alpha=-(3+\beta)/2$. The physical reason for the scale-free solution is thus that, contrary to the linear kernel, the behavior of the distribution near the cut-off points is unimportant for the (local) distribution. Other works have confirmed this power-law exponent, (\eg\ \citealt{white1982fss,1987JPhA...20L.801H}). If $\beta>1$ the integrals do not converge: \ie\ collisions with the boundaries of the distribution dominate and no power-law emerges. However, \citet{1975JAtS...32..380K} has shown that the $\beta=1$ kernel $K_{ij}=(m_i m_j)^{1/2}$, which therefore somewhat resembles the sum kernel, has a $\alpha=-2$ power-law.

\begin{figure*}
  \plottwo{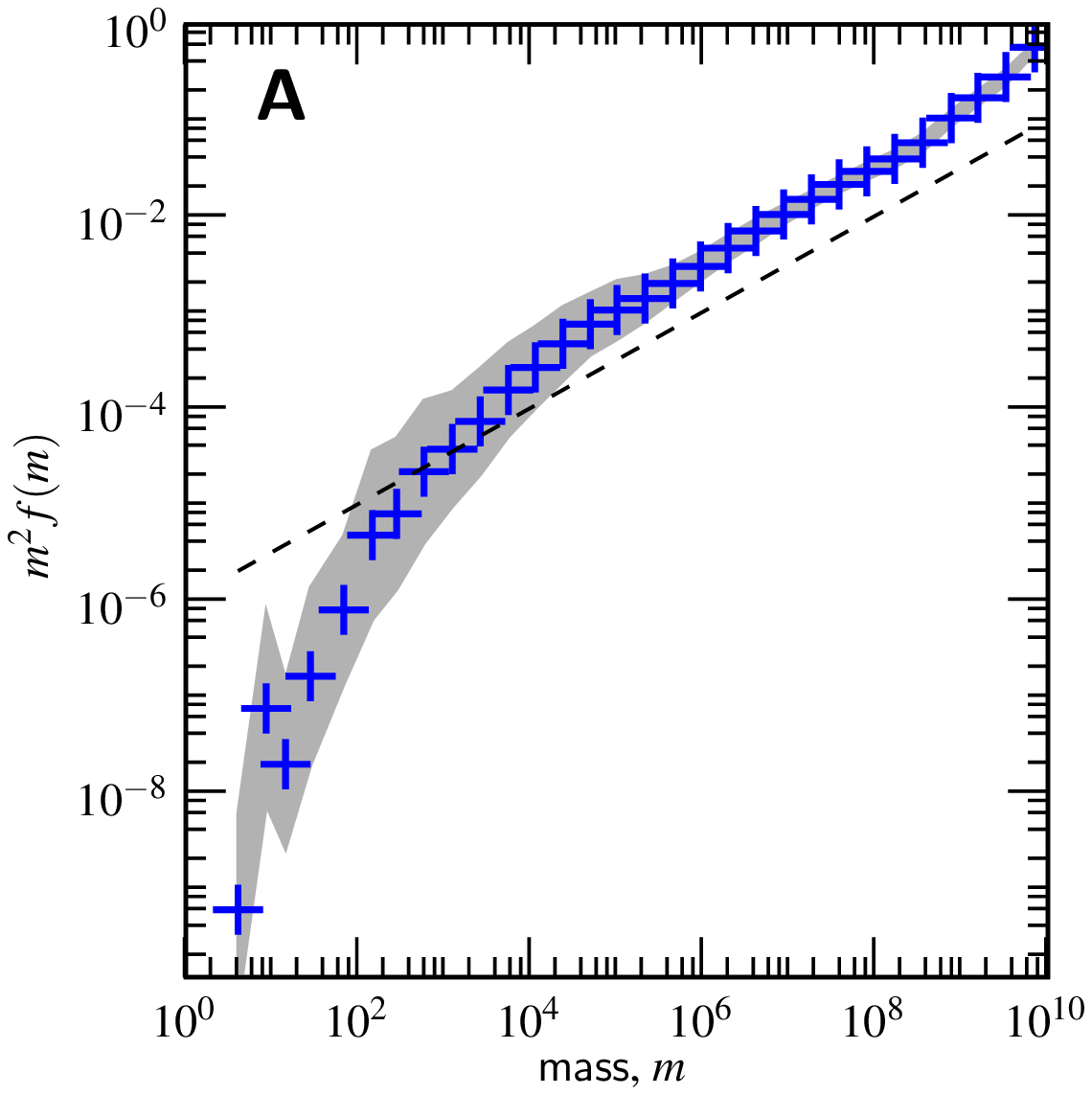}{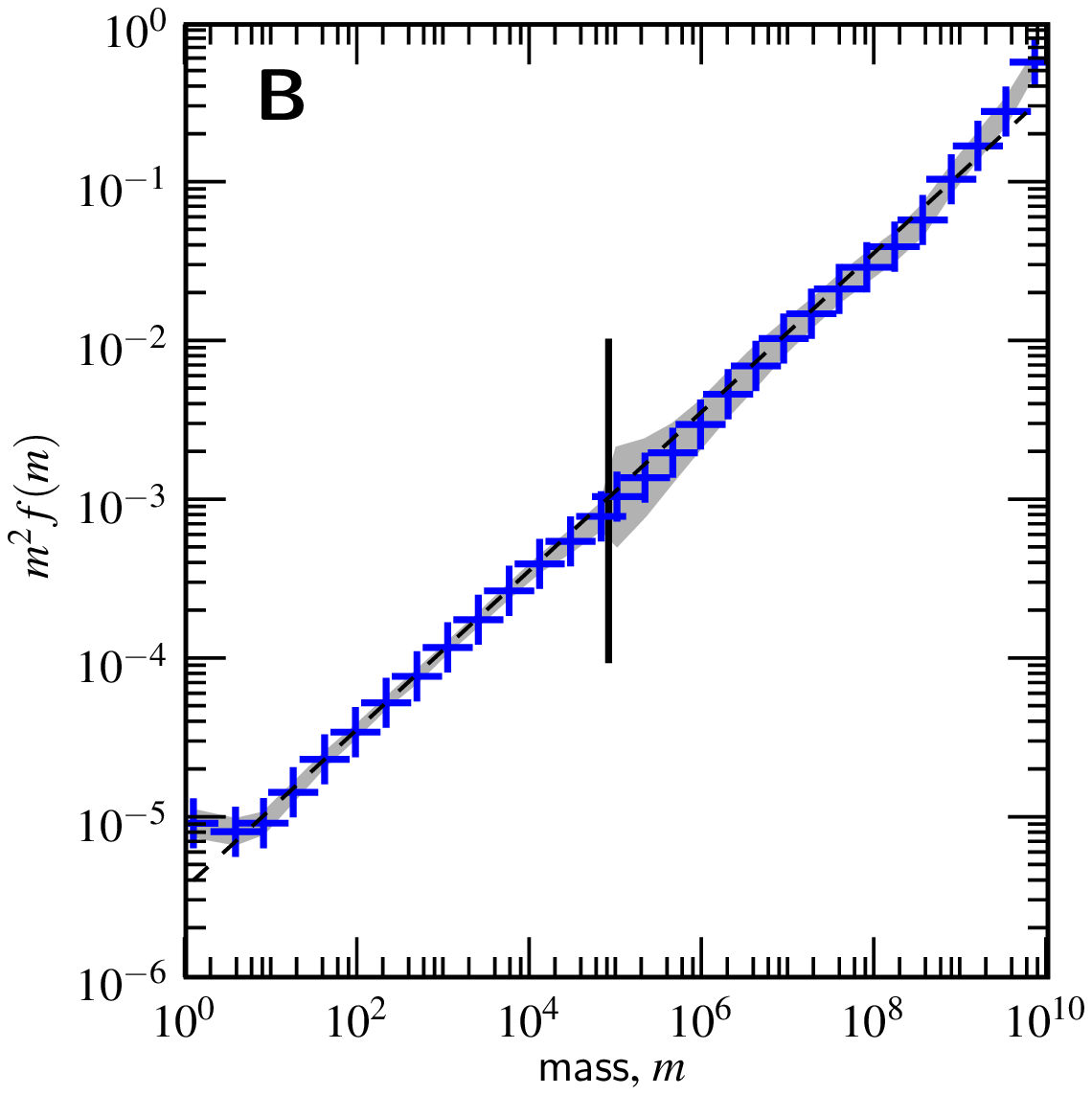}
  \caption{\label{fig:constantfrag}\textit{(A)} Catastrophic fragmentation in the constant kernel, $K=1$, with $m_\mathrm{frag}=10^{10}$. At masses above $m\sim10^4$ a power-law develops with a slope approximately equal to the predicted slope ($\alpha = -3/2$; dashed line). At low-$m$ the simulation lacks resolution and the power-law relation breaks down because of the episodic nature of the monomer injection. \textit{(B)} This discrepancy is solved by performing a follow-up simulation for the low-mass part of the distribution only ($m<m_\mathrm{cut}=10^5$) but including a removal rate due to collisions with large masses above $m_\mathrm{cut}$, $R(m)$, and a monomer injection rate, $I$, that were obtained in (A). The distribution above $m_\mathrm{cut}$ is copied from \textit{(A)}.}
\end{figure*}
To test whether a power-law emerges, we have performed fragmentation simulations for the constant kernel, $K=1$, see \fg{constantfrag}. Thus, as $\beta=0$ we expect to end up with a slope of $1/2$ for the $m^2f(m)$ mass distribution. This trend is indeed observed in \fg{constantfrag}a but only at large masses. The reason why an extension of the power-law to low masses is not seen, is the same as in \fg{sum}a. Again, due to the equal-mass method, region of low mass density are not represented. Additionally, the injection rate at $m=m_0$ is episodic: at one instance many monomers of total mass $\sim m_\mathrm{frag}$ enter the system, such that the mass density at this particular instance becomes very high (in comparison to the expected power-law distribution) for a very brief period of time. However, as coagulation is a non-linear process, \ie\ $dn_0/dt \propto -n_0^2$, these particles coagulate very quickly with each other. In other words, the episodic injection rate has the effect of speeding up the coagulation compared to a smooth rate and, when averaged over time, the true distribution is underestimated at places where the particle density strongly fluctuates with time. 

However, this deficiency of the equal-mass method does not affect the results at higher $m$: here the distribution is well resolved. Additionally, the mean (time-averaged) influx rate of monomers, $I$, at $m=m_0$ is available, as well as the removal rate of the higher mass particles, $R(m)$. A simple remedy to retrieve the low-$m$ distribution becomes then available. In a separate simulation we have focused only on the low-$m$ part of the mass spectrum until a cut-off mass $m_\mathrm{cut}$, and additionally include the fixed rates $I$ and $R$, which were obtained from the previous simulation. Thus, $R(m)$ is the probability that a particle of mass $m$ collides with \textit{any} particle of the `frozen' large-$m$ distribution that was previously obtained in \fg{constantfrag}a. Finally, if particles manage to grow above $m_\mathrm{cut}$ (\ie\ avoid to collide with a high-$m$ particle through $R$) they are also removed. In this way we effectively re-perform the same fragmentation simulation, but in a reduced simulation volume, where the much smaller inter-collision time-increments $\Delta t$ are appropriate for the low-$m$ tail. In \fg{constantfrag}b this procedure is illustrated. Here, we put the cut at $m\approx 10^5$, recalculated the steady-state for small $m$, and `glued' it to the high-$m$ distribution of $m>m_\mathrm{cut}$ from \fg{constantfrag}a.

\section{Summary and discussion}
\label{sec:summ}
\subsection{Merits of the Method}
In the previous section we have applied the grouping method to a variety of situations and found satisfactory behavior, both in terms of accuracy as in required computing power, despite the large dynamic range under consideration. The collisional evolutions of the analytic kernels were tested and agreed with the theoretical distributions. We also confirmed the predictions made for a class of runaway kernels, which required us to model an unprecedented large number of particles. Finally, we applied the grouping method to situations in which fragmentation operates and found excellent agreement with earlier findings (at least for the constant kernel; we are not aware of a prediction for the shape of the distribution function in the sum-fragmentation kernel).

Here we further discuss the (dis)advantages of both the equal mass and distribution methods and compare them with other studies. In short our conclusions are the following: the equal mass method is a true MC method, in the sense that no merging of the species array is required, while the mass density of the distribution is resolved (following a characteristic mass).  For non-runaway situations this is therefore the preferred method.  Other work along these lines can be found in \citet{2007physics...1103S}.  These authors implement a superparticle method to model rain drop formation.  These super-droplets (like our groups) represent a multiple number of physical particles with the same properties (like mass) and position.  However, their collisional setup is different. Where we allow many particles of one group to collide with another, in \citet{2007physics...1103S} there is at most one collision per particle.  Also, in the \citet{2007physics...1103S} work the number of superdroplets is conserved and therefore (if we understand their picture correctly) the mean mass of the superdroplets. Thus, they always need a large number of superdroplets in order to resolve the high-$m$ fluctuations (see the discussion on Fig.\ 2c of \citealt{2007physics...1103S}). The biggest difference, therefore, is that our method allows a dynamical adjustment of the group sizes. Thus, we think the \citet{2007physics...1103S} work is not capable of achieving the high dynamic range of our grouping method.  Still, we do note that their method achieves growth factors of at least $10^6$ in mass, and this may be sufficient for application to the atmospheric conditions of rain formation.

With the distribution method we are able to trace well the fluctuations of the mass (or distribution) spectrum, by choosing the zoom factors such that there are a fixed number of groups for all exponentially spaced mass bins. This method is therefore well suited for runaway kernels.  However, the consequence is a rapid proliferation of the $N_s$ parameter as many new species (which may be insignificant by, e.g., mass) are created. To limit $N_s$, these species need to be merged, which somewhat violates the spirit of MC-methods as this includes averaging, not only over mass but also over structural parameters.  The merging feature renders this method similar to Wetherill's discrete `batches' technique \citep{1990Icar...88..336W}.  However, in our method structural parameters, like porosity and charge, can be included and there is considerable (useful) freedom in how the averaging can be performed. These structural parameters survive, in averaged form, the merging process in our distribution method, and can be fundamental in the process of coagulation. For example, fractal grain structure (porosity) and dust charging can lead to, respectively, strong particle growth and even gelation \citep{2005NJPh....7..227K,2007A&A...461..215O}.

Both the \citet{1990Icar...88..336W} method and our approach give the correct behavior for the most massive and second most massive particles in the case of the product kernel (see \citealt{1999EP&S...51..205I} for verification).  Still, Wetherill's method has to deal with a number of free parameters: the spacing between the discrete batches and a parameter that determines the timestep, which must be fine-tuned. Our MC-method, although not free of them either, does have the advantage that the free parameters correspond to physically-identifiable quantities, such as the mass-fraction parameter, $f_\epsilon$.  Another key advantage of our code is that the timestep, $\Delta t$, follows naturally from the MC-code and is not required to be specified a priori. Especially in the super-runaway kernels with $\nu =2,3$, growth can be very erratic: during the runaway-process there is a large amount of growth during a small time step.  The success of Wetherill's approach lies in `discretizing' the original (mean-field) Smoluchowski equation in an efficient way.  Naturally, in Monte Carlo codes, discreteness is assured.

The existence of two separate methods might appear to be a drawback. That is, runaway kernels (modeled with the distribution method) and information on structural parameters (through the equal mass method) are difficult to treat simultaneously, especially because one does not know a priori if a runaway situation will materialize. However, information about the kernel is 'hard-wired' into the Monte Carlo program and it is therefore clear at which point the situation is susceptible to runaway growth. Then, these particles (the high-$m$ particles) have to be further resolved, and this can be achieved by switching to the distribution method. Thus, we feel that it is possible to model runaway kernels with MC-methods, provided one implements a hybride switch that signals runaway growth through the effective mass scaling exponent that a kernel exhibits during the simulation.  For example, consider the cross section of \eq{gravfoc}. Obviously, the mass scaling of $\sigma$, $m^\nu$, becomes stronger than unity ($\nu\approx 4/3$) for sufficiently large masses. It does so in a smooth manner, allowing a hybride switch to be implemented. That is, as the program signals a large value of $\nu$ for a particular kernel, it switches from the equal mass to the distribution method, \ie\ assigning lower zoom factors for the high-$m$ particles and, if needed,  merging structural parameters.  This kernel adaptation technique is analogous to adaptive mesh refinement \citep{2007ApJ...665..899W}, where the latter is grid based/spatial and we take a particle/stochastic approach.  Both numerical treatments require one to look carefully at the consequences of the switch, i.e., the impact of the merging prescription or the under-resolved part of the grid.

The bottom-line on MC-methods thus becomes clear: there is \textit{a priori} no universal way in which the correct, accurate result can be guaranteed. Each situation requires interpretation and, if necessary, (re)adjustment of some aspects related to the grouping method; most notably through the choices for the grouping factors, \ie\ the $\{z_i\}$. However, MC-methods are flexible enough to do so in physically meaningful way. As such, they are a very powerful tool to describe any dynamical system in which particles interact in a manner that can be statistically represented. In fact, the techniques presented in this work allow one to treat a wide range of, astrophysically relevant, kernels. The nicest feature of MC-methods is that besides mass, structural properties of particles, like porosity and charge, can be directly incorporated into the collisional dynamics.

\subsection{Astrophysical implications}
There are several astrophysical situations where a large dynamic range is required and for which our algorithm may be applicable. In protoplanetary disks, one of the key questions is how the tiny (sub)micron-sized grains are transferred to planetesimals \citep{2007prpl.conf..783D}. Because of the  increasing relative velocities with particle growth, it may very well be the case that, while grains grow, the likelihood of fragmentation increases. An additional argument for fragmentation is that collisional growth very efficiently removes the smallest grains, contrary to what observations indicate \citep{2005A&A...434..971D}. Recently, \citet{2007arXiv0711.2192B} has solved the steady-state coagulation-fragmentation equation for the dust distribution. It would be worthwhile to investigate whether these results are also obtained with the grouping method, and how this picture changes by inclusion of structural parameters, \eg\ the porosity of aggregates \citep{1993A&A...280..617O,1999Icar..141..388K,2007A&A...461..215O}.

If particles reach the planetesimal size ($\sim 1\ \mathrm{km}$) self-gravity may be sufficient to keep the collision products together. Furthermore, gravitational focussing, enhancing the gravitational cross-section, will turn the accretion process into a runaway process ($\nu>1$). \citet{2004ARA&A..42..549G} has outlined under which conditions runaway growth proceeds.  \citet{1978Icar...35....1G} followed the collisional evolution (including fragmentation) of a population of initially $10^{12}$ km-sized planetesimals until sizes of $\sim 500\ \mathrm{km}$ were reached (after which particle-in-a-box approximations fail). We again note that a bin-based approach may have difficulty to model the runaway stage unless it uses an ad-hoc extension to include the runaway particle explicitly \citep{1990Icar...88..336W}. \citet{2006AJ....131.2737B} have developed a hybrid code that combines a N-body and statistical approach. However, these models can only treat one variable; with a Monte Carlo approach the internal structure of the particles can be included and a rubble pile structure can be distinguished from a homogeneous one.

In stellar clusters, too, conditions may be feasible for runaway growth. Due to dynamical friction (equipartition) the more massive stars will sink to the center of the cluster and we may even have $\nu=2$ \citep{2001Icar..150..314M,2000Icar..143...74L}. Moreover, the cluster will eject the smaller members and contract as a whole. These processes lead to \textit{core collapse} in which the inner cut-off for the power law density function, $r_0$, disappears, $r_0\rightarrow 0$, and the density seemingly becomes singular. This condition is very favourable for a runaway-growth scenario, and may well be the mechanism for the formation of high mass stars, seed black holes or even gamma ray bursts \citep{1999A&A...348..117P,2006MNRAS.368..121F,2007Natur.450..388P}. Globular clusters are already successfully modeled by MC-methods where the structural parameters are now the energy and angular momentum characterizing an orbit on which a particle (group) resides \citep[e.g.][]{2000ApJ...540..969J}. Here, the grouping mechanism may then be well suited to model systems with a very large number of stars and stellar ejecta. Similarly, the collapse, and possible fragmentation, of an interstellar gas cloud, leading to the formation of a stellar cluster or black hole, could be treated by following many (porous) gas clumps.

\acknowledgements 
We appreciate the efforts of Volker Ossenkopf to proof-read the manuscript and the useful suggestions and comments of the two referees, which significantly improved the structure and presentation of this work.
\bibliographystyle{apj}
\bibliography{ms}
\appendix
\section{Grouped collision rates}
\label{app:app1}

We justify the expressions for the grouped collision rates, \eq{cr2}, where the collision rates are divided by the number of collisions during the group collisions. We show that, provided the individual rates $C_{ij}$ do not change as particles accrete, the mean collision rate is independent of the grouping.

Consider a species of $N$ identical particles that react with an individual rate $a_0$ with an external particle. In the language of the grouping method of \se{group} the $N$ particles are of species $j$ that, during the group process, react with a particle from species $i$ and $a_0=C_{ij}$. Let the process be such that the $N$ particles react in groups of $N/w$, where $1 \le w \le N$ is the total number of such events.  

Let $P_i$ be the probability that the system is in state $i$, meaning that the $i$-th event (out of $w$) has occurred. At $t=0$ the system is in the first state, so $P_0(t=0)=1$ and $P_i(t=0)=0$ for $i\neq 1$. Let the rate at which the system evolves from state $i$ to the next state ($i+1$) be given by $\lambda_i$, then

\begin{eqnarray}
    \frac{dP_0}{dt} = - \lambda_0 P_0;\\
    \frac{dP_1}{dt} = - \lambda_1 P_1 + \lambda_0 P_0; \\
    \frac{dP_i}{dt} = - \lambda_i P_i + \lambda_{i-1} P_{i-1}; \qquad  (1<i<w)\\
    \frac{dP_w}{dt} = + \lambda_{w-1} P_{w-1}.
  \label{eq:state}
\end{eqnarray}
Let $\overline{R(t)}$ be the mean reactivity of the system measuring the mean number of particles that has reacted at time $t$ ($0\le R \le 1$). Likewise, let $\overline{R^2}$ measure the square number of reactants, \ie
\begin{equation}
  \overline{R^k}(t) = \frac{1}{N^k} \sum_{i=0}^{w} P_i \left( \frac{iN}{w} \right)^k = \sum_{i=0}^w P_i \left( \frac{i}{w} \right)^k.
\label{eq:relmoments}
\end{equation}
If we weigh the state-equations (\eqp{state}) with $i/n$ and $(i/n)^2$, respectively, and sum over $i$ we obtain:
\begin{eqnarray}
  \nonumber
  \frac{d\overline{R}}{dt} = \frac{1}{w} \sum_{i=0}^{w} \lambda_i P_i;\\
  \frac{d\overline{R^2}}{dt} = \frac{1}{w^2} \sum_{i=0}^w (2i+1) \lambda_i P_i.
  \label{eq:moments}
\end{eqnarray}
Now the $\lambda_i$ are proportional to the amount of particles left, \ie\ $\lambda_i \propto (1 - i/w)N$. In the case of group collisions, furthermore, we have argued that the rates should be divided by the number of reactants, such that the collision rate stays the same. We thus obtain
\begin{equation}
  \lambda_i = (w-i) a_0,
  \label{eq:lambda}
\end{equation}
where $a_0$ is the individual collision rate between two single particles. \Eq{moments} then transform to
\begin{eqnarray}
  \frac{d\overline{R}}{dt} = a_0 \left( 1 - \overline{R} \right);\\
  \frac{d\overline{R^2}}{dt} = a_0 \left( -2\overline{R^2} + \frac{2w-1}{w}\overline{R} + \frac{1}{w} \right).
\end{eqnarray}
This first equation of the first moment (average value) is independent of $w$. Its solution is of course the well known exponential decay corresponding to Poisson statistics, \ie
\begin{equation}
  \overline{R}(t) = 1-\exp[-at].
  \label{eq:Rt}
\end{equation}
Inserting $\overline{R}$ into the equation for the second moment we solve for $\overline{R^2}$ as
\begin{equation}
  \overline{R^2}(t) = \left( 1-e^{-at} \right)\left( 1-e^{-at}\left( 1-w^{-1} \right) \right).
\label{eq:moment2}
\end{equation}
The spread $\sigma$ in the distribution then becomes
\begin{equation}
  \sigma^2(w) = \left( 1 - e^{-at} \right) \frac{e^{-at}}{w}.
\label{eq:charmass}
\end{equation}
That is, for $w\rightarrow \infty$, $\sigma\rightarrow 0$ and we are certain about the fraction of objects that have collided. For small $w$, however, there can be a significant spread.

By choosing the collision rates as in \eq{lambda} -- which is simply the rate for a collision with one particle, $a_0(N-iN/w)$, divided by the number of particles in the group ($N/w$) -- we end up with \eq{Rt} for the mean number of collisions. Since this is independent of $w$ the mean reactivity of the system is also unaffected by the choice of $w$. For the mean reactivity it does not matter whether $w=N$ (no grouping) or $w=1$ (all particles in the same group) as long as $a_0$ stays constant during the group process. However, the spread in $\overline{R}$ does depend on the choice of $w$, which is quite natural, of course. A similar procedure cannot be performed for the in-group collisions since here \eq{state} involves $i^2$ terms. The group collision rates for in-group collisions are not exact. However, the rates of \eq{cr2} are still a good approximation, and will become exact for large $w$. Besides, in-group collisions are a relatively minor occurrence.
\end{document}